\documentclass[prd,onecolumn]{revtex4}
\usepackage{dcolumn}
\usepackage{multirow}
\usepackage{graphicx}
\usepackage{amssymb}
\usepackage{bm}
\usepackage{hyperref}
\usepackage{epstopdf}
\usepackage{color}
\usepackage{mathrsfs}
\usepackage{amsmath,amssymb,amsthm}
\begin{document}
\title{Exploring the deviation of cosmological constant by a generalized pressure dark energy model}
\author{Jun-Chao Wang$^{1}$}
\email{dakaijun@dakaijun.cn}
\author{Xin-He Meng$^{1}$}
\email{xhm@nankai.edu.cn}
\affiliation{
$^1${Department of Physics, Nankai University, Tianjin 300071, China}\\}
\begin{abstract}
We bring forward a generalized pressure dark energy (GPDE) model to explore the evolution of the universe. This model has covered three common pressure parameterization types and can be reconstructed as quintessence and phantom scalar fields, respectively. We adopt the cosmic chronometer (CC) datasets to constrain the parameters.  The results show that the inferred late-universe parameters of the GPDE model are (within $1\sigma$): The present value of Hubble constant $H_{0}=(72.30^{+1.26}_{-1.37})$km s$^{-1}$ Mpc$^{-1}$; Matter density parameter $\Omega_{\text{m0}}=0.302^{+0.046}_{-0.047}$, and the universe bias towards quintessence. While when we combine CC data and the $H_0$ data from Planck, the constraint implies that our model matches the $\Lambda$CDM model nicely. Then we perform dynamic analysis on the GPDE model and find that there is an attractor or a saddle point in the system corresponding to the different values of parameters. Finally, we discuss the ultimate fate of the universe under the phantom scenario in the GPDE model. It is demonstrated that three cases of pseudo rip, little rip, and big rip are all possible.
\end{abstract}
\maketitle
\section{Introduction} \label{sec1}
Over the past two decades, a large number of cosmological observations has confirmed that the expansion of the late universe is speeding up \cite{riess1998observational,perlmutter1999constraining,eisenstein2005detection,bennett2013nine,ade2014planck}, which has become one of the greatest challenges of cosmology. In order to explain the accelerated expansion, there are two main approaches: modifying gravity (MG) and adding dark energy (DE). The former means modifies the geometric parts of General relativity (GR), such as scalar-tensor theory \cite{dvali20004d}, $f(R)$ gravity \cite{carroll2004cosmic} and brane cosmology \cite{brans1961mach,sahni2003braneworld}. The other method is to add the dark energy which breaks the strong energy condition and produces a mysterious repulsive force to make the universe accelerate the expansion. The simplest DE model is the $\Lambda$CDM model, where the cosmological constant $\Lambda$ is related to DE, and its equation of state (EoS) is $\omega_{\text{de}}= -1$. The $\Lambda$CDM model provides a fairly good explanation for current cosmic observations. Recently, Planck-2018 reaffirmed the validity of the 6-parameter $\Lambda$CDM model in describing the evolution of the universe \cite{aghanim2018planck}. Nonetheless, there are two long-term problems with the $\Lambda$CDM model. One is the fine-tuning problem: The observation of dark energy density is 120 orders of magnitude smaller than the theoretical value in quantum field theory \cite{carroll2001cosmological,weinberg1989cosmological}; The second is the coincidence problem: At the beginning of the universe, the proportion of DE is especially tiny while now the ratio of the dark energy density and matter density are exactly on the same magnitude. Besides, in recent years, the tension of Hubble constant between the Planck datasets and SHoES has reached to 4.4$\sigma$ \cite{riess2019Large}, which are under the $\Lambda$CDM model and the cosmic distance ladder, respectively.  So as to alleviate these problems, many dynamic dark energy models with the time-variation EoS are proposed, including scalar field models (such as quintessence \cite{1998cosmological,kamenshchik2001alternative,amendola2000coupled,chiba2000kinetically,zlatev1999quintessence}, phantom \cite{caldwell2002phantom,2003phantom,nojiri2003quantum}, k-essence \cite{armendariz2001essentials,deffayet2011k,scherrer2004purely}, quintom \cite{cai2010quintom,guo2005cosmological} and tachyon \cite{bagla2003cosmology}), holographic model \cite{li2004model}, agegraphic model \cite{wei2008new}, chaplygin gas model \cite{bento2002generalized,gorini2003can} and so on. 

The model presented in this paper is also a dynamic dark energy model, which parameterizes the total pressure of the universe. Parameterization of the observable is an effective method to explore the characteristics of DE, such as the parameterization of EoS \cite{chevallier2001accelerating,linder2003exploring,barboza2009generalized}, luminosity distance \cite{cattoen2007hubble,gruber2014cosmographic}, dark energy density \cite{alam2004there}, pressure \cite{sen2008deviation,kumar2013deviation,zhang2015exploring,yang2016diagnostics,wang2017new,wang2018pressure} and deceleration factors \cite{akarsu2012cosmological}. Taking the pressure parameterization as an example. In general, we can write the pressure parameter equation as $P=\sum_{n=0} P_{n}x_{n}(z)$, where $x_{n}(z)$ expands at the late universe as the following forms (i) Redshift: $x_{n}=z^n$, (ii) Scale factor: $x_n(z)=(1-a)^n=(z/(1+z))^n$, (iii) Logarithmic form: $x_n(z)=(\ln (1+z))^n$. The form corresponding to $n=1$ in (i) and (ii) was proposed by Zhang, Yang, Zou, et al.\ \cite{zhang2015exploring}. Case (iii) for $n=1$ was given by Wang and Meng \cite{wang2018pressure}. In order to unify these mainstream parameterization methods, we suggest a three-parameter pressure parameterization model to explore the evolution of the universe.

The content of this paper is organized as follows: Sec.\ \ref{sec2} presents a generalized pressure dark energy (GPDE) model of the total pressure and discusses its feature. In Sec.\ \ref{sec3}, we use CC datasets to impose constraints on the parameters of the GPDE model. The discussion of fixed points under the GPDE model is analyzed in Sec.\ \ref{sec4}. In Sec.\ \ref{sec5}, we exhibit the end of the universe under the phantom case. The last section Sec.\ \ref{sec6} is the conclusion.
\section{THEORETICAL MODEL} \label{sec2}
Pressure parameterization describes our universe in the following ways: First, hypothesize a relationship between the pressure $P$ and the redshift $z$. Then the expression of the density $\rho$ can be derived from the conservation equation $\dot{\rho}+3(\dot{a}/a)(\rho +P)=0$. Finally, by utilizing the Friedmann equations $H^2=3/(8\pi G)\sum_i \rho_i$ and the EoS $\omega=P/\rho$, we can get the form of the Hubble parameter $H$ and $\omega$, respectively. Here we take the speed of light as $c=1$. At this point, a closed system of cosmic evolution has been established which is described by the Friedmann equations, the conservation equation, and the EoS form. 
It is worth noting that there are still some deviations between the $\Lambda$CDM and actual (e.g., $H_0$ tension), but the physical mechanism behind it is not clear. Taking advantage of this kind of handwritten model, we can probe the possible deviations further between the dynamic case and the cosmological constant case without a specific premise. In this work, we propose a generalized pressure dark energy model of the total energy components in a spatially flat Fridenmann-Robertson-Walker (FRW) universe
\begin{equation}
P(z)=P_{1}-P_{2}\left[\frac{(1+z)^{-\beta}-1}{\beta}\right],\quad \beta\neq 0,
\label{1}
\end{equation}
Where $P_1$, $P_2$ and $\beta$ are free parameters. Notice that $P_1$ is the current value of the total pressure in the universe, and $P_2$ represents the deviation of $P(z)-z$. The model degenerates into the $\Lambda$CDM model as $P_2=0$. To mention, this parametric form of $P(z)$, i.e.\ Eq.\ (\ref{1}), is inspired by a generalized equation of state for dark energy \cite{barboza2009generalized}. When specific limits are given to $\beta$, this model returns to the three models mentioned in Sec.\ \ref{sec1}, i.e.
\begin{equation}
\label{2}
P(z) = \left\{\begin{array}{ll} P_{a}+P_{b}z,&$\text{for}$\ \beta= -1 \\\\
P_a +P_{b}\ln(1+z),\ \ \ \ \ \ \ \ \ \ &$\text{for}$\
\beta\rightarrow 0 \\\\
P_{a}+P_{b}{\left(\frac{z}{1+z}\right)},&$\text{for}$\ \beta=+1
\end{array}\right..
\end{equation}
By using Eq.\ (\ref{1}), the relationship of scale factor $a$ ($a=1/(1+z)$) and the conservation equation ($\dot{\rho}+3(\dot{a}/a)(\rho +P)=0$), we can get the density as
\begin{equation}
\label{3}
\rho(a)=\frac{3a^\beta P_2}{(3+\beta)\beta}-\frac{P_2}{\beta}-P_1+C a^{-3},
\end{equation}
Where $C$ is the integral constant. We assume $\rho_0$ is the current total density, i.e.\ $\rho(a = 1)= \rho_0$. Finally, the total density and total pressure can be respectively sorted into the following form
\begin{equation}
\label{4}
\rho(a)=\rho_0\left( P_a a^{\beta}+\Omega_{\text{m0}}a^{-3}+1-P_a-\Omega_{\text{m0}}\right),
\end{equation}
\begin{equation}
\label{5}
P(a)=\rho_0\left[-(1-P_a -\Omega_{\text{m0}})-\frac{3+\beta}{3}P_a a^{\beta}\right].
\end{equation}
Parameters $P_1$ and $P_2$ have been replaced here by new parameters $P_a$ and $\Omega_{\text{m0}}$, where $P_a\equiv 3P_2 /((3+\beta)\beta\rho_0)$, $\Omega_{\text{m0}}\equiv (\beta \rho_{0}+P_2 +\beta P_1 -3P_2/(3+\beta))/(\beta \rho_0)$. In the density expression (\ref{4}), the item $\rho_{0}\Omega_{\text{m0}}a^{-3}$ is corresponding to the matter density $\rho_{m}$. So $\Omega_m|_{a=1}=\rho_{m}/\rho=\Omega_{\text{m0}}$ signifies that the physical meaning of the parameter $\Omega_{\text{m0}}$ is the present-day matter density parameter. The term $\rho_{0}(P_aa^{\beta}+1-P_a-\Omega_{\text{m0}})$ accords with the dark energy density $\rho_{\text{de}}$, and the constant part $1-P_a-\Omega_{\text{m0}}$ looks similar to the $\Lambda$CDM case. The term $P_aa^{\beta}$ makes $\rho_{\text{de}}$ change with time: The larger the $|P_a|$ is, the more deviation from the $\Lambda$CDM model will be; The larger the $\beta$ is, the faster the dark energy density will change. Accordingly, this cosmological model only includes matter and dark energy components, and the pressure of dark energy $P_{\text{de}}$ is the total pressure $P$. From $(\rho_{\text{de}}/\rho_{m})|_{a\rightarrow 0,\beta>-3} \rightarrow 0$ we can also know that for the case of $\beta>-3$, the DE accounts for a small percentage in the early universe. Note that when $\beta = -3$, the density of the part of the dark energy is expressed as the matter density, and the total density $\rho(a)$ of our GPDE model is equivalent to the $\Lambda$CDM model.

Suppose the dark energy is a scalar field $\phi$ that changes with time. The corresponding pressure and density are equivalent to $\rho_{\text{de}} = (n/2)\dot{\phi}^2+V(\phi)$ and $P_{\text{de}}=(n/2)\dot{\phi}^2-V(\phi)$, separately, where $n = 1$ or $-1$ corresponds to the quintessence and  phantom scalar field, respectively. The calculation shows that $\dot{\phi}^2=-(n\beta /3)\rho_0P_aa^{\beta}$, so $\beta P_a<0$ fits quintessence, and $\beta P_a> 0$ matches for phantom.

Additionally, for the GPDE model, the EoS of the dark energy $\omega_{\text{de}}$, the dimensionless Hubble parameter $E$, the deceleration parameter $q$ and the jerk parameter $j$ respectively take the form as
\begin{equation}
\label{6}
\omega_{\text{de}}=-1-\frac{\frac{1}{3}\beta P_a a^{\beta}}{1-P_a-\Omega_{\text{m0}}+P_a a^{\beta}},
\end{equation}
\begin{equation}
\label{7}
E=\left(\Omega_{\text{m0}}a^{-3}+P_a a^{\beta}+1-P_a-\Omega_{\text{m0}}\right)^{1/2},
\end{equation}
\begin{equation}
\label{8}
q\equiv -\frac{\ddot{a}}{aH^2}=-1+\frac{3}{2}\Omega_m-\frac{\beta P_a}{2E^2}a^{\beta},
\end{equation}
\begin{equation}
\label{9}
j\equiv\frac{\dddot{a}}{aH^3}=1+2(q+1)(2q-1)-\frac{a^{\beta}P_{a}\beta ((\beta +3)}{2E^2}.
\end{equation}

\section{RESULTS OF THE DATA ANALYSIS} \label{sec3}
By measuring the age difference between two galaxies under different redshifts, we can get the Hubble constant $H(z)$, called cosmic chronometer data. In this section, We constrain our parameter by 33 unrelated cosmic chronometer data listed in table \ref{t1}, spanning the redshift range $0< z <2$. The optimal values of the parameters can be obtained by taking the minimum value of $\chi ^2$, which is expressed as
\begin{equation}
\label{10}
\chi^{2}=\sum_{i=1}^{33}\left[\frac{H_{\text{obs}}(z_{i})-H_{0}E(z_{i}) }{\sigma_{i}^{2}}\right] ^{2},
\end{equation}
with the corresponding four-dimensional parameter space $\{ H_0, \Omega_{\text{m0}}, P_a, \beta \}.$
\begin{table}[ht]
\centering
\begin{tabular}{llll|llll|llll}
\hline\hline
$z\quad$    & $H(z)\quad$ & $\sigma_{H(z)}\quad$ & Ref.$\quad$& $z\quad$   & $H(z)\quad$ & $\sigma_{H(z)}\quad$ & Ref.$\quad$ & $z\quad$   & $H(z)\quad$ & $\sigma_{H(z)}\quad$ & Ref.$\quad$\\
\hline
0.07     & 69 & 19.68  & \cite{zhang2014four}  & 0.36 & 81.2 & 5.9 & \cite{moresco2012improved} & 0.7812 & 105 & 12 & \cite{moresco2012improved}\\
0.09     & 69 & 12  & \cite{jimenez2003constraints}    & 0.3802 & 83 & 13.5 & \cite{moresco20166} & 0.8754 & 125 & 17 & \cite{moresco2012improved}\\
0.1      & 69 & 12 & \cite{stern2010d}  & 0.4 & 95 & 17 & \cite{simon2005constraints} & 0.88 & 90 & 40 & \cite{stern2010d}    \\
0.12      & 68.6 & 26.2   & \cite{zhang2014four}     & 0.4004 & 77 & 10.2 & \cite{moresco20166} & 0.9 & 117 & 23 & \cite{simon2005constraints}\\
0.17   & 83  & 8   & \cite{simon2005constraints} & 0.4247 & 87.1 & 11.2 & \cite{moresco20166} & 1.037 & 154 & 20 & \cite{moresco2012improved}\\
0.1791   & 75  & 4   & \cite{moresco2012improved} & 0.4497  & 92.8 & 12.9 &\cite{moresco20166} & 1.3 & 168 & 17 & \cite{simon2005constraints}\\
0.1993       & 75  & 5 & \cite{moresco2012improved}   & 0.47 & 89 & 50 & \cite{wang2017clustering} & 1.363 & 160 & 33.6 & \cite{moresco2015raising}\\
0.2     & 72.9  & 29.6 & \cite{zhang2014four}      & 0.4783 & 80.9 & 9 & \cite{moresco20166} & 1.43 & 177 & 18 & \cite{simon2005constraints}\\
0.27     & 77 & 14 & \cite{simon2005constraints}  & 0.48 & 97 & 62 & \cite{stern2010d} & 1.53 & 140 & 14 & \cite{simon2005constraints}\\
0.28   & 88.8 & 36.6 & \cite{zhang2014four} & 0.5929 & 104 & 13 & \cite{moresco2012improved} & 1.75 & 202 & 40 & \cite{simon2005constraints}\\
0.3519 & 83.0 & 14 & \cite{moresco2012improved}  & 0.6769 & 92 & 8 & \cite{moresco2012improved} & 1.965 & 186.5 & 50.4 & \cite{moresco2015raising} \\
\hline\hline
\end{tabular}
\caption{Cosmic chronometers data.}
\label{t1}
\end{table}

\begin{figure}
\centering
\includegraphics[scale=1]{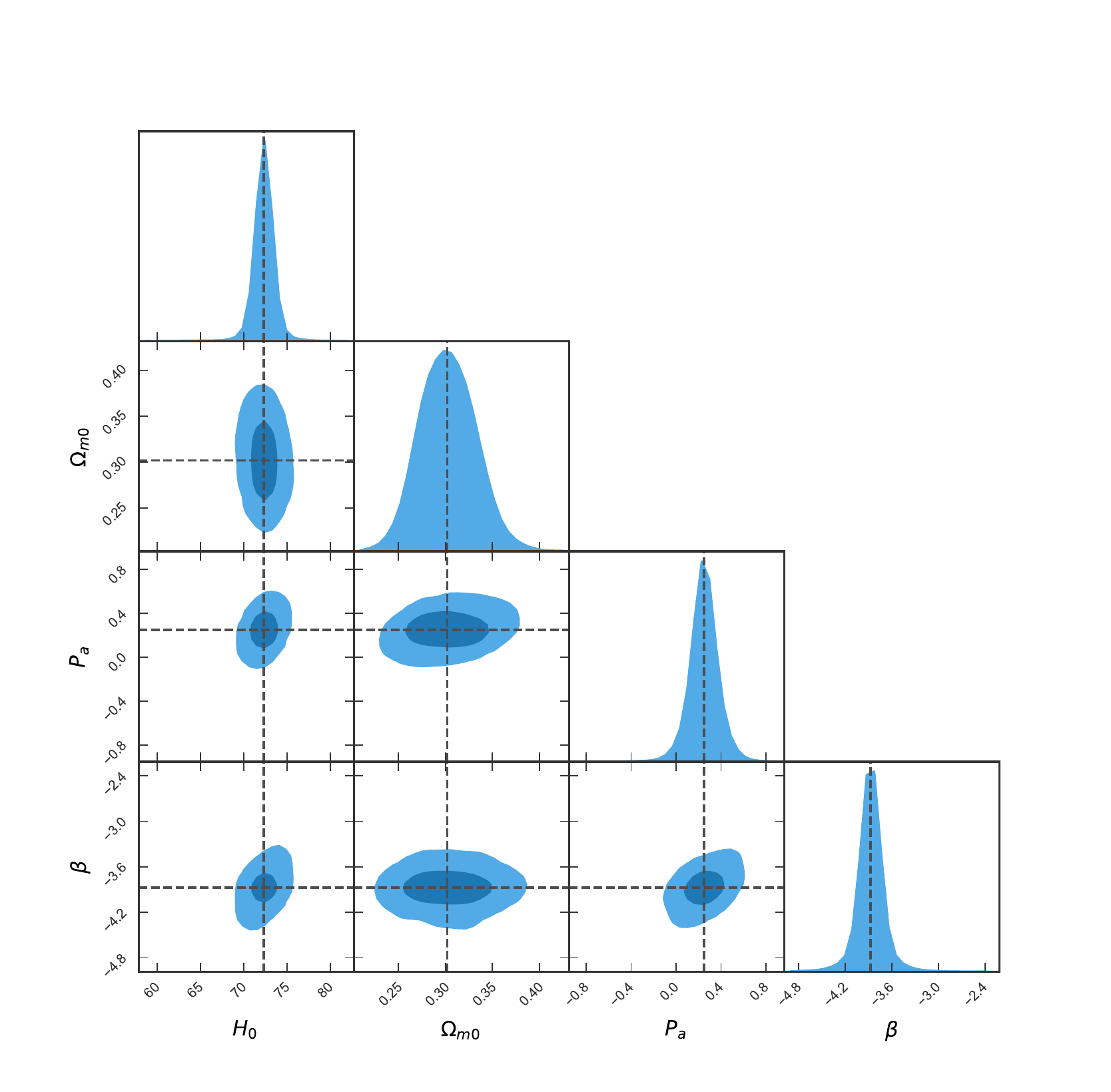}
\caption{the 1-dimensional and 2-dimensional marginalized probability distributions of the GPED model by using CC observations. The dark blue and light blue areas represent $1\sigma$ and $2\sigma$ errors, respectively.}
\label{f4}
\end{figure}

\begin{figure}
\centering
\includegraphics[scale=0.43]{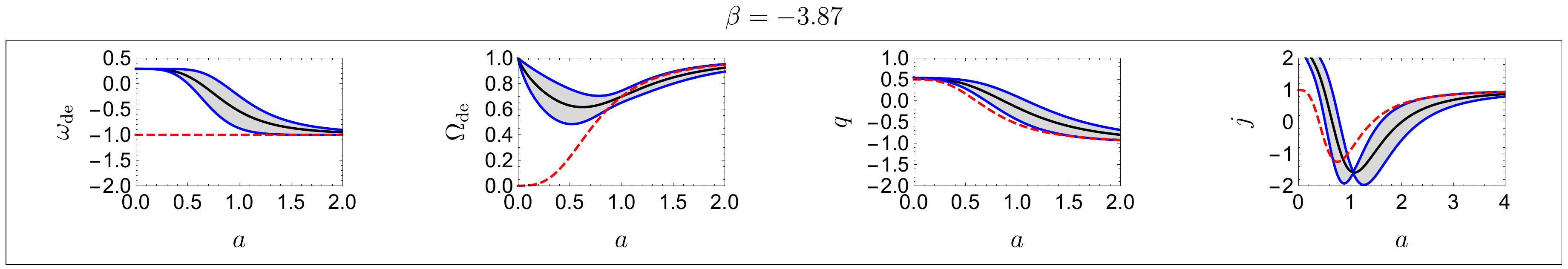}
\caption{The evolutions of $\omega_{\text{de}}, \Omega_{\text{de}}, q$ and $j$ for $\beta=-3.87$. In each panel, The black line and red dashed line correspond to the GPDE model and the $\Lambda$CDM model, respectively, with the best-fit values listed in Table \ref{t5}. The shaded region and blue lines represent the $1\sigma$ level regions and corresponding boundaries in the GPDE model.}
\label{f5}
\end{figure}

As the first attempt, we adopt the Monte Carlo Markov chain (MCMC) method and use the python package emcee \cite{foreman2013emcee} to produce a MCMC sample with CC data. The results are displayed as a contour map by another python package pygtc \cite{Bocquet2016}. We list the priors and initial seeds on the parameter space in Table \ref{t4}.
\begin{table}[ht]
\centering
\begin{tabular}{ccc}
\hline\hline
Parameter & Prior & Initial seed\\
\hline
$H_0$ & $[0,100]$& 69\\
$\Omega_{\text{m0}}$ & $[0,1]$ & 0.33\\
$P_a$ & $[-1,1]$ & $0$ \\
$\beta$ &$[-5,5]$&$-0.02$\\
\hline\hline
\end{tabular}
\caption{The priors and initial seeds of parameters used in the pasterior analysis.}
\label{t4}
\end{table}
Figure \ref{f4} shows the 1-dimensional and 2-dimensional marginalized probability distributions of the GPED model.
In the meantime，the best-fit values and $1\sigma$ confidence level for the $H_0$, $\Omega_{\text{m0}}$, $P_a$ and $\beta$ are listed in Table \ref{t5}.
\begin{table}[ht]
\centering
\begin{tabular}{cc}
\hline\hline
Parameter\quad & Best-fit value with 1$\sigma$ error\\
\hline
$H_0$  & $72.30^{+1.26}_{-1.37}$\\
$\Omega_{\text{m0}}$ &$0.302^{+0.046}_{-0.047}$\\
$P_a$ & $0.249^{+0.160}_{-0.180}$ \\
$\beta$ &$-3.87^{+1.77}_{-1.66}$\\
\hline
$\chi ^2_{min}$ & $14.4053$\\
\hline\hline
\end{tabular}
\caption{Best-fit parameters $H_0$, $\Omega_{\text{m0}}$,$P_a$ and $\beta$ from CC datasets. Here we also list $\chi ^2_{min}$. }
\label{t5}
\end{table}
From the constraint results, $P_{a} \beta <0$ in 1$\sigma$, which indicates our universe is under quintessence situation and has some deviation from the $\Lambda$CDM model in a point of view of data.  The differences of $H_0$ between our results and SHoES \cite{riess2019Large} and Planck base-$\Lambda$CDM \cite{aghanim2018planck} are 0.9 $\sigma$ and 3.5 $\sigma$, respectively. The evolution of DE EoS parameter $\omega_{\text{de}}$, DE density parameter $\Omega_{\text{de}}$, deceleration parameter $q$ and jerk parameter $j$ with 1 $\sigma$ error propagation from data fitting (Table \ref{t5}) are shown in figure \ref{f5}. 

\begin{figure}
\centering
\includegraphics[scale=0.7]{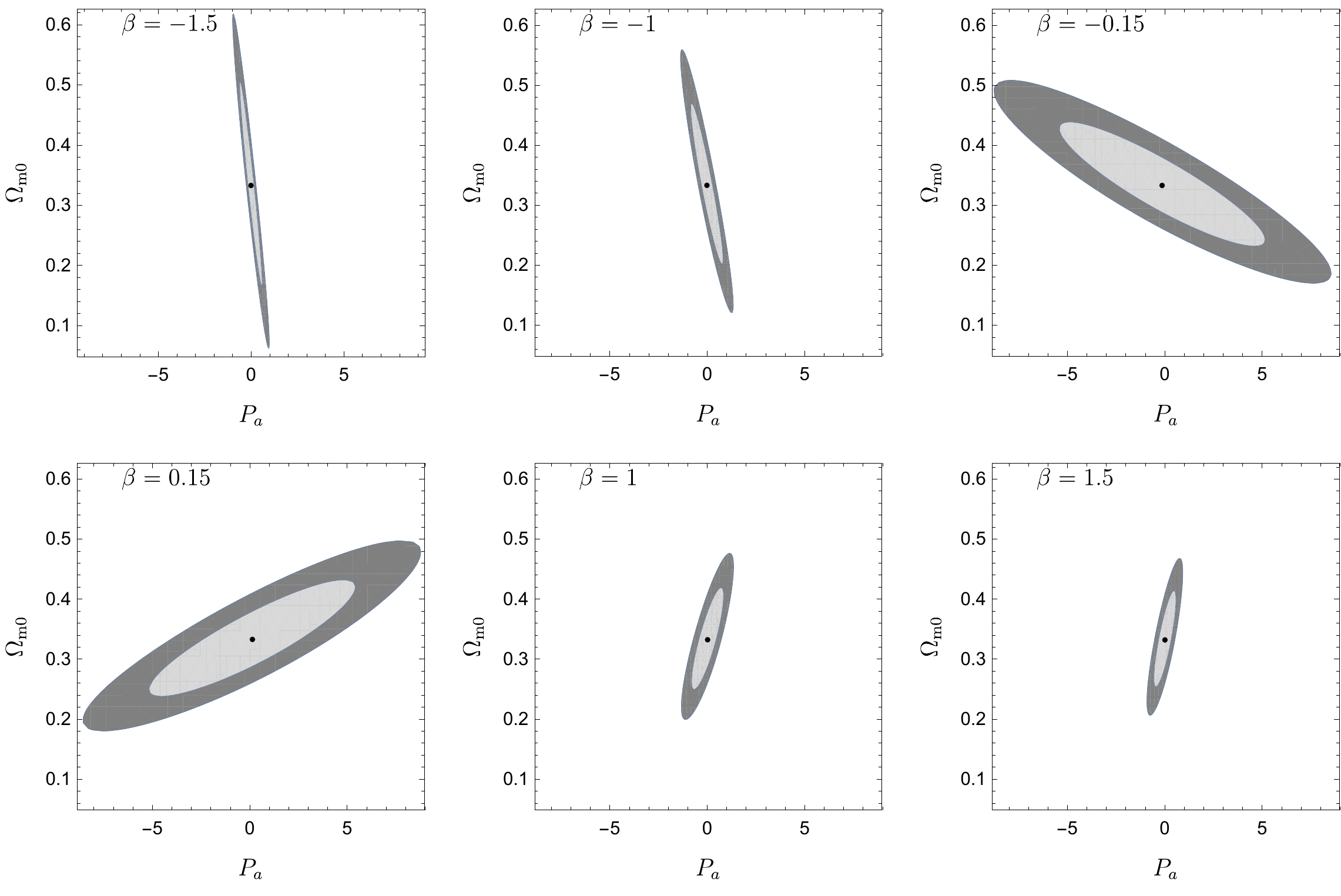}
\caption{The $1\sigma$ and $2\sigma$ contours in the planes of $P_a - \Omega_{\text{m0}}$ for different choices of $\beta$ at present by using CC datasets and the $H_0$ data. In each panel, the black dot represents the best-fit values of $P_a - \Omega_{\text{m0}}$.}
\label{f1}
\end{figure}

\begin{figure}
\includegraphics[scale=0.43]{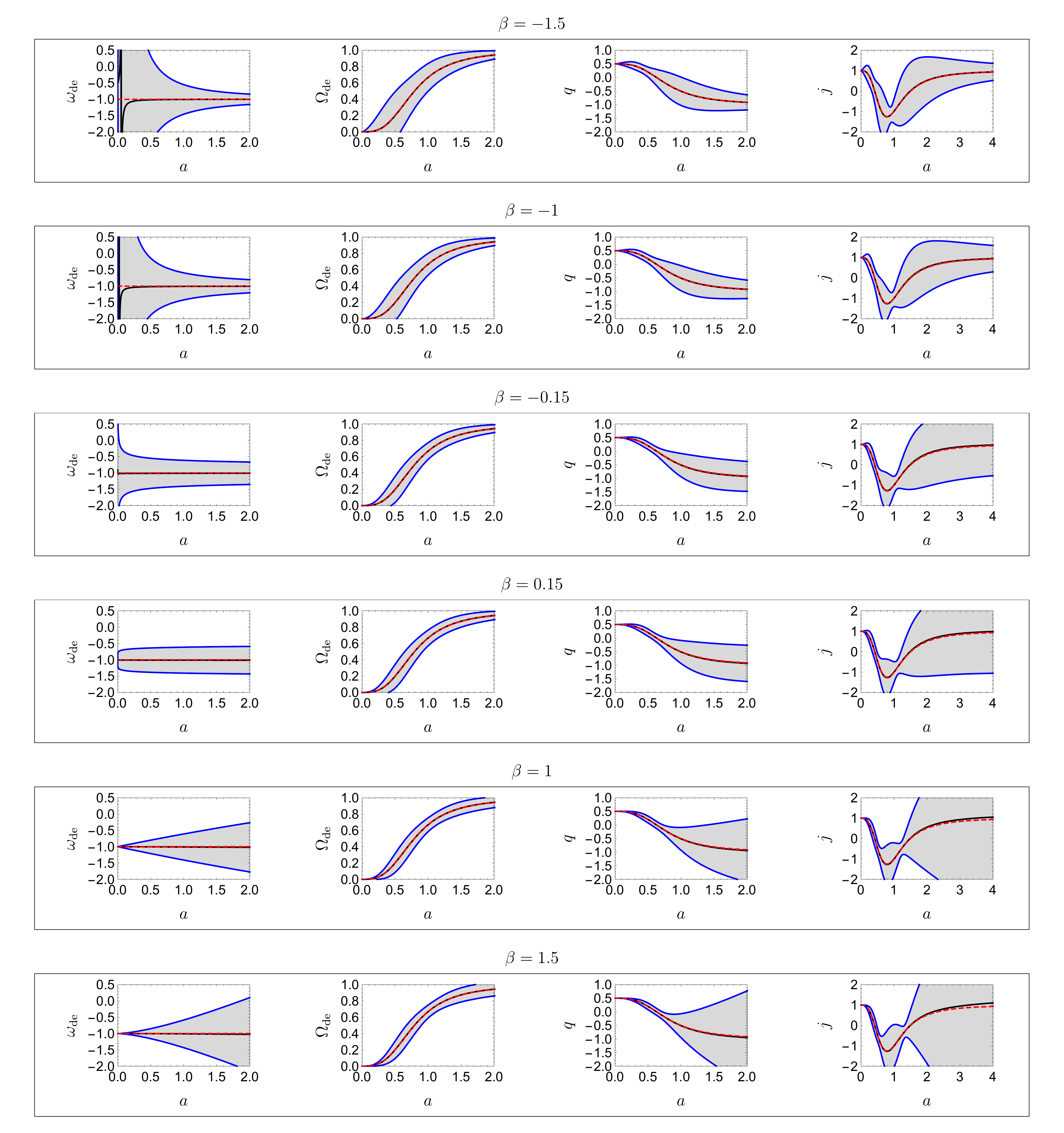}
\caption{The evolution of $\omega_{\text{de}}, \Omega_{\text{de}}, q$ and $j$ for different choices of $\beta$. In each panel, The black line and red dashed line correspond to the GPDE model and the $\Lambda$CDM model, respectively, with the best-fit values listed in Table \ref{t2}. The shaded region and blue lines represent the $1\sigma$ level regions and corresponding boundaries in the GPDE model.}
\label{f2}
\end{figure}

From figure \ref{f5} we find that these results are acceptable, except for the DE density parameter $\Omega_{\text{de}}$, which is too high at the beginning of the universe and contradicts the facts we now know. For this reason, we make a further try: while using CC data, we also combine the data of Planck-2018 \cite{aghanim2018planck} with the Hubble constant $H_0=67.4$km s$^{-1}$ Mpc$^{-1}$ and pick $\beta=-1.5, -1, -0.15, 0.15, 1$ and $1.5$, respectively to constrain the parameter pair $\{P_a$, $\Omega_{\text{m0}}\}$. The results are shown in figure \ref{f1}. Table \ref{t2} lists the best-fit values and $1\sigma$ confidence level for the $P_a$ and $\Omega_{\text{m0}}$ under CC data and the $H_0$ data joint constraints. We discover that for the six kinds of circumstances, the best values of $\Omega_{\text{m0}}$ are all around $0.33$, and $\beta P_a>0$ indicates that the universe is slightly biased toward phantom but still includes quintessence within $1\sigma$ confidence level. The value of $P_a$ is small, meaning in this case the deviation of this model from the $\Lambda$CDM model is not significant. The minimum $\chi^2$ of these six cases are very close, implying that this model is not very sensitive to the selected values of $\beta$, that is, the degeneracy is high. Figure \ref{f2} shows the difference in export parameters $\omega_{\text{de}}, \Omega_{\text{de}}, q$ and $j$ between the GPDE model and the $\Lambda$CDM model with the best-fit values for the considered values of $\beta$. It can be concluded from figure \ref{f2} that the distinction between these two models is almost indistinguishable. The exceptions are for the cases of $\beta=-1.5$ and $-1$, whose $\omega_{\text{de}}$ rapidly increase and decrease at the beginning of the universe, and then stabilize at a position slightly less than $-1$.  Unlike the first attempt, these fitting results show that our model is consistent with the $\Lambda$CDM model. The reason is estimated to be that in the second try, the data of $H_0$ is from Planck, which depends on the $\Lambda$CDM model.

\begin{table}[ht]
\centering
\begin{tabular}{l|llllll}
\hline\hline
$\beta$ & $-1.5$ & $-1$ & $-0.15$ & $0.15$ & $1$ & $1.5$\\
\hline
$P_a$ & $-8.47$e$-3$\quad\quad & $-1.71$e$-2$\quad\quad & $-1.35$e$-1$\quad\quad & $1.37$e$-1$\quad\quad & $-1.93$e$-2$\quad\quad & $1.15$e$-2$\\
$1\sigma$ & $0.6\sim 0.7$ & $0.8\sim 0.9$ & $5.2\sim 5.4$ & $5.2\sim 5.4$\quad\quad & $0.8\sim 0.9$ & $0.5\sim 0.6$\\
$\Omega_{\text{m0}}$ & $0.333$ & $0.333$ & $0.333$ &$0.333$ &$0.332$ &$0.332$ \\
$1\sigma$ &$0.107$&$0.133$&$0.103$&$0.097$&$0.085$&$0.080$\\
$\chi_{min}^2$&$14.6368$&$14.6363$&$14.6358$&$14.6357$&$14.6360$&$14.6363$\\
\hline\hline
\end{tabular}
\caption{Best-fit parameters $P_a$ and $\Omega_{\text{m0}}$ from CC datasets (in combination with the local value of the Hubble parameter $H_0$) for different choices of $\beta$. Here we also quote $\chi_{min}^2$ in every case of $\beta$. Line 3 and line 5 are $1\sigma$ errors of $P_a$ and $\Omega_{\text{m0}}$, respectively.}
\label{t2}
\end{table}
\section{DYNAMIC ANALYSIS} \label{sec4}
\begin{figure}
\includegraphics[scale=0.6]{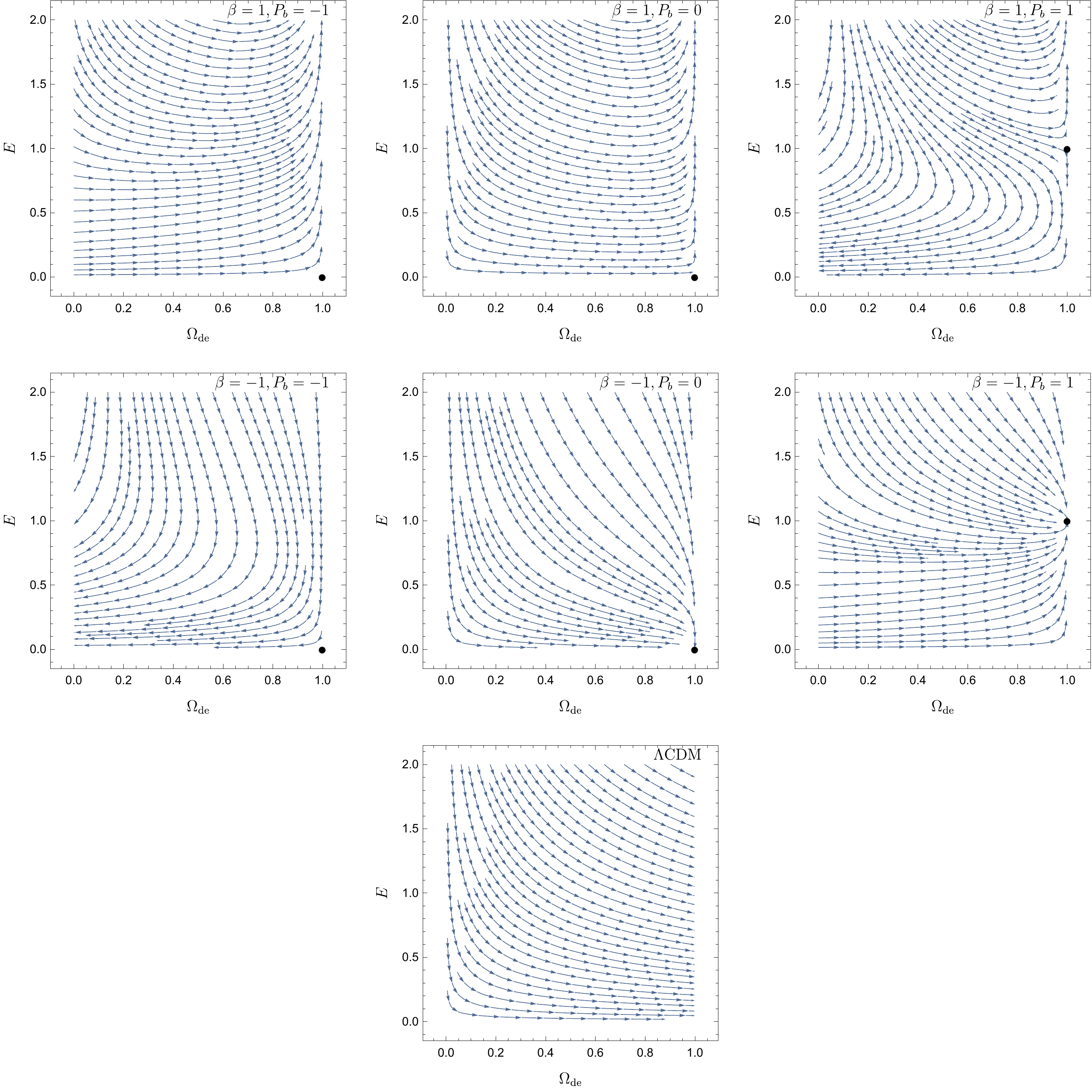}
\caption{The dynamic system evolution of the GPDE model. The black dots represent the fixed point. For comparison, we also exhibit the scenario of the $\Lambda$CDM model in the last row. The arrows stand for the direction of time.}
\label{f3}
\end{figure}
In this section, we will construct a self-consistent dynamical system to analyze the cosmological evolution of the GPDE model. Select $\Omega_{\text{de}}$ and $E$ as independent variables. By Friedmann equation, we can get the self-consistent dynamic system as 
\begin{equation}
\label{11}
\Omega_{\text{de}}'=(1-\Omega_{\text{de}})\left(\beta\Omega_{\text{de}}-\frac{P_b \beta}{E^{2}}+3\Omega_{\text{de}}\right),
\end{equation}
\begin{equation}
\label{12}
E'=\frac{E}{2}\left[\beta\Omega_{\text{de}}-\frac{P_b\beta}{E^{2}}-3(1-\Omega_{\text{de}})\right].
\end{equation}
where $P_b\equiv 1-P_a-\Omega_{\text{m0}}$, "$'$" represents the derivative of $\ln (a)$. The following are common methods for finding fixed points and its stability of a system. Let $\Omega_{\text{de}}'=E'=0$, then for $P_b\geq 0$, there is a fixed point $\{\Omega_{\text{de}}=1, E=P_b^{1/2}\}$ in the dark energy dominant period. Do the perturbation expansion of the system near the fixed point and then we can get the Jacobian matrix
\begin{equation}
\label{13}
M\equiv\left( \begin{matrix}
     \frac{\partial\Omega_{\text{de}}'}{\Omega_{\text{de}}} & \frac{\partial\Omega_{\text{de}}'}{\partial E} \\
    \frac{\partial E'}{\partial\Omega_{\text{de}}} & \frac{\partial E'}{\partial E}
\end{matrix}\right)
=\left( \begin{matrix}
     -3 & 0 \\
    \frac{\beta +3}{2}P_b ^{1/2} &\beta
\end{matrix}\right) 
\end{equation}
Its eigenvalues are $\lambda_1=-3$ and $\lambda_2=\beta$. The stability of the system depends on the sign of the eigenvalues $\lambda_1$ and $\lambda_2$. For the GPDE model, $\beta$ is a real number and not equal to zero, so there are two situations: When $\beta<0$, point $\{1,P_b^{1/2}\}$ is an attractor, and when $\beta>0$, point $\{1,P_b^{1/2}\}$ is a saddle point. While for $P_b<0$, we can determine the properties of the fixed point by observing its figure. On the other side, for the case of $\Lambda$CDM, the parameter $P_a=0$, so $\beta \Omega_{\text{de}}-\frac{P_{b}\beta}{E^2} = \beta\frac{P_a a^{\beta}}{E^2} =0$, the dynamic system goes to 
\begin{equation}
\label{14}
\Omega_{\text{de}}'=3\Omega_{\text{de}}(1-\Omega_{\text{de}}),
\end{equation}
\begin{equation}
\label{15}
E'=-\frac{3}{2}E(1-\Omega_{\text{de}}).
\end{equation}
Figure \ref{f3} illustrates the evolution of the dynamics system for $\beta = -1$, $1$ and $P_b = -1$, $0$, $1$. There also exhibits the evolutional trajectories of the $\Lambda$CDM model as a comparison in the last row. From figure \ref{f3}, it can be found that there is a saddle point $\{1,0\}$ for $P_b<0$. When $\beta>0$ and $P_b<0$, it shows that $E\rightarrow\infty$ and $\Omega_{\text{de}}\rightarrow 1$, which is the case of phantom. While for $\beta<0$ and $P_b<0$, one can see $\Omega_{\text{de}}\rightarrow 0$  which corresponds to quintessence. When $P_b\geq0$, whether the universe is biased towards quintessence or phantom depends on $\beta$ and initial conditions which cannot be judged directly. Based on the above analysis, we summarize the stability of the GPDE model in table \ref{t3}.
\begin{table}[ht]
\centering
\begin{tabular}{l|l|l}
\hline\hline
 & $\{\Omega_{\text{de}},E\}$ & Stability \\
\hline
$P_b\geq 0$,$\beta<0$ & $\{1,P_b^{1/2}\}$ &  attractor \\
$P_b\geq 0$,$\beta>0$ & $\{1,P_b^{1/2}\}$ &  saddle point \\
$P_b< 0$& $\{1,0\}$ &  saddle point \\
\hline\hline
\end{tabular}
\caption{The stability of the GPDE model.}
\label{t3}
\end{table}
\section{universe fate under the phantom field} \label{sec5}
In recent years, various data have shown that the dark energy state equation is close to $−1$: the WMAP9 have found $\omega=-1.073^{+0.090}_{-0.089}$ based on WMAP+CMB+BAO+$H_0$ \cite{bennett2013nine}. Planck-2018 constrained the EoS as $\omega=-1.03\pm 0.03$ with SNe \cite{aghanim2018planck}. Moreover DES suggests $\omega=-0.978\pm 0.059$ in the joint analysis of DES-SN3YR+CMB \cite{abbott2019first}. Using existing data, it is still impossible to distinguish between phantom ($\omega<-1$) and non-phantom($\omega \geq -1$), that is, the phantom case cannot be excluded. Phantom has gradually increased energy density, and eventually, the universe accelerates so fast that the particles lost contact with each other and rip apart. Based on the various evolutionary behaviors of $H(t)$, the final fate of the universe can be divided into the following three categories \cite{frampton2012pseudo}:
\begin{itemize}
\item[1] Big rip: $H(t)\rightarrow\infty$ as $t \rightarrow$ constant, so the rip will happen at a certain time.
\item[2] Little rip: $H(t)\rightarrow\infty$ as $t \rightarrow \infty$. This situation has no singularities in the future.
\item[3] Pseudo rip: $H(t)\rightarrow \text{constant}$. This is the case with the de-Sitter universe and little rip.
\end{itemize}

In this section, we are interested in the rip case of our GPED model. For the GPED model, Hubble constant is written as
\begin{equation}
\label{16}
H=\frac{\dot{a}}{a}=H_0\left(\Omega_{\text{m0}}a^{-3}+P_a a^{\beta}+1-P_a-\Omega_{\text{m0}}\right)^{1/2}.
\end{equation}
In section \ref{sec2}, we have pointed out that $\beta P_a>0$ for phantom case of the the GPED model. Next we discuss each situation separately.
\begin{itemize}
\item[1] $\beta<0, P_a< 0$:
\end{itemize}

With the growth of cosmic time, $a\rightarrow\infty$ and 
\begin{equation}
\label{17}
H\rightarrow H_0(1-P_a-\Omega_{\text{m0}})^{1/2},
\end{equation}
i.e.\ $H(t)$ tends to be a constant, and the universe will approach the de-Sitter universe infinitely. So this case is pseudo rip.
\begin{itemize}
\item[2] $\beta>0, P_a> 0$:
\end{itemize}

When $a\rightarrow\infty$, 
\begin{equation}
\label{18}
H\rightarrow H_0(P_a a^{\beta})^{1/2}.
\end{equation}
By solving the above differential equation, we can get the relation between scale factor $a$ and time $t$, which can be written as
\begin{equation}
\label{19}
a=\left( 1-\frac{\beta P_a^{1/2}H_0}{2}(t-t_0)\right)^{-2/\beta},
\end{equation}
Where $t_0$ is the present value of time. Substitute Eq.\ (\ref{19}) for Eq.\ (\ref{18}), one obtains
\begin{equation}
\label{20}
H=\frac{H_0P_a^{1/2}}{1-\frac{\beta P_a^{1/2}H_0}{2}(t-t_0)}.
\end{equation}
So when $t\rightarrow \frac{2}{\beta P_a^{1/2}H_0}+t_0$, $H(t)\rightarrow \infty$ which means the universe will have a big rip after $t-t_0=\frac{2}{\beta P_a^{1/2}H_0}$. Thus, the lifetime of the universe is determined by two parameters, $\beta$ and $P_a$, regardless of the matter density $\Omega_{\text{m0}}$. Let us make a rough estimate. Take $\beta=1$, $P_a=0.02$, $H_0=70$km s$^{-1}$ Mpc$^{-1}$, then the universe will be torn apart after 198Gyr. If $13.8$Gyr is the current age of the universe, then the universe has spent only 6.5\% of its life.
\begin{itemize}
\item[3] $\beta\rightarrow 0$:
\end{itemize}

In this case, the GPDE model degenerates into the model in \cite{wang2018pressure}. According to the discussion in \cite{wang2018pressure}, the ultimate fate of the universe is the little rip.

To sum up, there are three possible fates under the phantom of the GPDE model:
\begin{itemize}
\item[1] Big rip for $\beta>0, P_a > 0$;
\item[2] Little rip for $\beta\rightarrow 0$;
\item[3] Pseudo rip for $\beta<0, P_a<0$.
\end{itemize}

\section{conclusions} \label{sec6}
In principle, it is interesting to insert models or theories into a more general framework to test their validity. Not only does this reveal a new set of solutions, but it may also enable more accurate consistency checks for the original model. This paper has made this attempt by expanding the $\Lambda$CDM model to a generalized pressure dark energy (GPDE) model. The GPED model has three independent parameters: The present value of matter density parameter $\Omega_{\text{m0}}$, the parameter $P_a$ which represents the deviation from the $\Lambda$CDM model, and the parameter $\beta$. Picking different values of parameter $\beta$ can produce three common pressure parametric models. By using the cosmic chronometer (CC) datasets to constrain parameters, it shows that Hubble constant is $H_{0}=(72.30^{+1.26}_{-1.37})$km s$^{-1}$ Mpc$^{-1}$. And the differences of $H_0$ between our results and SHoES \cite{riess2019Large} and Planck base-$\Lambda$CDM \cite{aghanim2018planck} are 0.9 $\sigma$ and 3.5 $\sigma$, respectively. In addition, for the GPDE model, the matter density parameter is $\Omega_{\text{m0}}=0.302^{+0.046}_{-0.047}$, and the universe bias towards quintessence in $1\sigma$ error. While when we combine CC datasets and the $H_0$ data from Planck, the constraint implies that our model matches the $\Lambda$CDM model well. Then we explore the fixed point of this model and find that there is an attractor or a saddle point corresponding to the different values of parameters. Next, we analyze the rip of the universe under phantom case and draw the conclusion that there are three possible endings of the universe: Pseudo rip for $\beta <0$, $P_a<0$, big rip for $\beta >0$, $P_a>0$ and little rip for $\beta\rightarrow 0$. Finally, we estimate that for the big rip case, the universe has a life span of 198Gyr.

Dark energy has been proposed for twenty years, but its nature remains unknown. With this model, we can probe the possible deviation further between the dynamic case and the cosmological constant condition through existing data. 
\section*{ACKNOWLEDGEMENTS} \label{sec7}
Jun-Chao Wang thanks Wei Zhang for the helpful discussions and code guidance about MCMC.

\bibliographystyle{apsrev4-1}
\bibliography{bibfile}

\begin{thebibliography}{56}%
\makeatletter
\providecommand \@ifxundefined [1]{%
 \@ifx{#1\undefined}
}%
\providecommand \@ifnum [1]{%
 \ifnum #1\expandafter \@firstoftwo
 \else \expandafter \@secondoftwo
 \fi
}%
\providecommand \@ifx [1]{%
 \ifx #1\expandafter \@firstoftwo
 \else \expandafter \@secondoftwo
 \fi
}%
\providecommand \natexlab [1]{#1}%
\providecommand \enquote  [1]{``#1''}%
\providecommand \bibnamefont  [1]{#1}%
\providecommand \bibfnamefont [1]{#1}%
\providecommand \citenamefont [1]{#1}%
\providecommand \href@noop [0]{\@secondoftwo}%
\providecommand \href [0]{\begingroup \@sanitize@url \@href}%
\providecommand \@href[1]{\@@startlink{#1}\@@href}%
\providecommand \@@href[1]{\endgroup#1\@@endlink}%
\providecommand \@sanitize@url [0]{\catcode `\\12\catcode `\$12\catcode
  `\&12\catcode `\#12\catcode `\^12\catcode `\_12\catcode `\%12\relax}%
\providecommand \@@startlink[1]{}%
\providecommand \@@endlink[0]{}%
\providecommand \url  [0]{\begingroup\@sanitize@url \@url }%
\providecommand \@url [1]{\endgroup\@href {#1}{\urlprefix }}%
\providecommand \urlprefix  [0]{URL }%
\providecommand \Eprint [0]{\href }%
\providecommand \doibase [0]{http://dx.doi.org/}%
\providecommand \selectlanguage [0]{\@gobble}%
\providecommand \bibinfo  [0]{\@secondoftwo}%
\providecommand \bibfield  [0]{\@secondoftwo}%
\providecommand \translation [1]{[#1]}%
\providecommand \BibitemOpen [0]{}%
\providecommand \bibitemStop [0]{}%
\providecommand \bibitemNoStop [0]{.\EOS\space}%
\providecommand \EOS [0]{\spacefactor3000\relax}%
\providecommand \BibitemShut  [1]{\csname bibitem#1\endcsname}%
\let\auto@bib@innerbib\@empty
\bibitem [{\citenamefont {Riess}\ \emph {et~al.}(1998)\citenamefont {Riess},
  \citenamefont {Filippenko}, \citenamefont {Challis}, \citenamefont
  {Clocchiatti}, \citenamefont {Diercks}, \citenamefont {Garnavich},
  \citenamefont {Gilliland}, \citenamefont {Hogan}, \citenamefont {Jha},
  \citenamefont {Kirshner} \emph {et~al.}}]{riess1998observational}%
  \BibitemOpen
  \bibfield  {author} {\bibinfo {author} {\bibfnamefont {A.~G.}\ \bibnamefont
  {Riess}}, \bibinfo {author} {\bibfnamefont {A.~V.}\ \bibnamefont
  {Filippenko}}, \bibinfo {author} {\bibfnamefont {P.}~\bibnamefont {Challis}},
  \bibinfo {author} {\bibfnamefont {A.}~\bibnamefont {Clocchiatti}}, \bibinfo
  {author} {\bibfnamefont {A.}~\bibnamefont {Diercks}}, \bibinfo {author}
  {\bibfnamefont {P.~M.}\ \bibnamefont {Garnavich}}, \bibinfo {author}
  {\bibfnamefont {R.~L.}\ \bibnamefont {Gilliland}}, \bibinfo {author}
  {\bibfnamefont {C.~J.}\ \bibnamefont {Hogan}}, \bibinfo {author}
  {\bibfnamefont {S.}~\bibnamefont {Jha}}, \bibinfo {author} {\bibfnamefont
  {R.~P.}\ \bibnamefont {Kirshner}},  \emph {et~al.},\ }\href@noop {}
  {\bibfield  {journal} {\bibinfo  {journal} {The Astronomical Journal}\
  }\textbf {\bibinfo {volume} {116}},\ \bibinfo {pages} {1009} (\bibinfo {year}
  {1998})}\BibitemShut {NoStop}%
\bibitem [{\citenamefont {Perlmutter}\ \emph {et~al.}(1999)\citenamefont
  {Perlmutter}, \citenamefont {Turner},\ and\ \citenamefont
  {White}}]{perlmutter1999constraining}%
  \BibitemOpen
  \bibfield  {author} {\bibinfo {author} {\bibfnamefont {S.}~\bibnamefont
  {Perlmutter}}, \bibinfo {author} {\bibfnamefont {M.~S.}\ \bibnamefont
  {Turner}}, \ and\ \bibinfo {author} {\bibfnamefont {M.}~\bibnamefont
  {White}},\ }\href@noop {} {\bibfield  {journal} {\bibinfo  {journal}
  {Physical Review Letters}\ }\textbf {\bibinfo {volume} {83}},\ \bibinfo
  {pages} {670} (\bibinfo {year} {1999})}\BibitemShut {NoStop}%
\bibitem [{\citenamefont {Eisenstein}\ \emph {et~al.}(2005)\citenamefont
  {Eisenstein}, \citenamefont {Zehavi}, \citenamefont {Hogg}, \citenamefont
  {Scoccimarro}, \citenamefont {Blanton}, \citenamefont {Nichol}, \citenamefont
  {Scranton}, \citenamefont {Seo}, \citenamefont {Tegmark}, \citenamefont
  {Zheng} \emph {et~al.}}]{eisenstein2005detection}%
  \BibitemOpen
  \bibfield  {author} {\bibinfo {author} {\bibfnamefont {D.~J.}\ \bibnamefont
  {Eisenstein}}, \bibinfo {author} {\bibfnamefont {I.}~\bibnamefont {Zehavi}},
  \bibinfo {author} {\bibfnamefont {D.~W.}\ \bibnamefont {Hogg}}, \bibinfo
  {author} {\bibfnamefont {R.}~\bibnamefont {Scoccimarro}}, \bibinfo {author}
  {\bibfnamefont {M.~R.}\ \bibnamefont {Blanton}}, \bibinfo {author}
  {\bibfnamefont {R.~C.}\ \bibnamefont {Nichol}}, \bibinfo {author}
  {\bibfnamefont {R.}~\bibnamefont {Scranton}}, \bibinfo {author}
  {\bibfnamefont {H.-J.}\ \bibnamefont {Seo}}, \bibinfo {author} {\bibfnamefont
  {M.}~\bibnamefont {Tegmark}}, \bibinfo {author} {\bibfnamefont
  {Z.}~\bibnamefont {Zheng}},  \emph {et~al.},\ }\href@noop {} {\bibfield
  {journal} {\bibinfo  {journal} {The Astrophysical Journal}\ }\textbf
  {\bibinfo {volume} {633}},\ \bibinfo {pages} {560} (\bibinfo {year}
  {2005})}\BibitemShut {NoStop}%
\bibitem [{\citenamefont {Bennett}\ \emph {et~al.}(2013)\citenamefont
  {Bennett}, \citenamefont {Larson}, \citenamefont {Weiland}, \citenamefont
  {Jarosik}, \citenamefont {Hinshaw}, \citenamefont {Odegard}, \citenamefont
  {Smith}, \citenamefont {Hill}, \citenamefont {Gold}, \citenamefont {Halpern}
  \emph {et~al.}}]{bennett2013nine}%
  \BibitemOpen
  \bibfield  {author} {\bibinfo {author} {\bibfnamefont {C.~L.}\ \bibnamefont
  {Bennett}}, \bibinfo {author} {\bibfnamefont {D.}~\bibnamefont {Larson}},
  \bibinfo {author} {\bibfnamefont {J.}~\bibnamefont {Weiland}}, \bibinfo
  {author} {\bibfnamefont {N.}~\bibnamefont {Jarosik}}, \bibinfo {author}
  {\bibfnamefont {G.}~\bibnamefont {Hinshaw}}, \bibinfo {author} {\bibfnamefont
  {N.}~\bibnamefont {Odegard}}, \bibinfo {author} {\bibfnamefont
  {K.}~\bibnamefont {Smith}}, \bibinfo {author} {\bibfnamefont
  {R.}~\bibnamefont {Hill}}, \bibinfo {author} {\bibfnamefont {B.}~\bibnamefont
  {Gold}}, \bibinfo {author} {\bibfnamefont {M.}~\bibnamefont {Halpern}},
  \emph {et~al.},\ }\href@noop {} {\bibfield  {journal} {\bibinfo  {journal}
  {The Astrophysical Journal Supplement Series}\ }\textbf {\bibinfo {volume}
  {208}},\ \bibinfo {pages} {20} (\bibinfo {year} {2013})}\BibitemShut
  {NoStop}%
\bibitem [{\citenamefont {Ade}\ \emph {et~al.}(2014)\citenamefont {Ade},
  \citenamefont {Aghanim}, \citenamefont {Armitage-Caplan}, \citenamefont
  {Arnaud}, \citenamefont {Ashdown}, \citenamefont {Atrio-Barandela},
  \citenamefont {Aumont}, \citenamefont {Baccigalupi}, \citenamefont {Banday},
  \citenamefont {Barreiro} \emph {et~al.}}]{ade2014planck}%
  \BibitemOpen
  \bibfield  {author} {\bibinfo {author} {\bibfnamefont {P.~A.}\ \bibnamefont
  {Ade}}, \bibinfo {author} {\bibfnamefont {N.}~\bibnamefont {Aghanim}},
  \bibinfo {author} {\bibfnamefont {C.}~\bibnamefont {Armitage-Caplan}},
  \bibinfo {author} {\bibfnamefont {M.}~\bibnamefont {Arnaud}}, \bibinfo
  {author} {\bibfnamefont {M.}~\bibnamefont {Ashdown}}, \bibinfo {author}
  {\bibfnamefont {F.}~\bibnamefont {Atrio-Barandela}}, \bibinfo {author}
  {\bibfnamefont {J.}~\bibnamefont {Aumont}}, \bibinfo {author} {\bibfnamefont
  {C.}~\bibnamefont {Baccigalupi}}, \bibinfo {author} {\bibfnamefont {A.~J.}\
  \bibnamefont {Banday}}, \bibinfo {author} {\bibfnamefont {R.}~\bibnamefont
  {Barreiro}},  \emph {et~al.},\ }\href@noop {} {\bibfield  {journal} {\bibinfo
   {journal} {Astronomy \& Astrophysics}\ }\textbf {\bibinfo {volume} {571}},\
  \bibinfo {pages} {A16} (\bibinfo {year} {2014})}\BibitemShut {NoStop}%
\bibitem [{\citenamefont {Dvali}\ \emph {et~al.}(2000)\citenamefont {Dvali},
  \citenamefont {Gabadadze},\ and\ \citenamefont {Porrati}}]{dvali20004d}%
  \BibitemOpen
  \bibfield  {author} {\bibinfo {author} {\bibfnamefont {G.}~\bibnamefont
  {Dvali}}, \bibinfo {author} {\bibfnamefont {G.}~\bibnamefont {Gabadadze}}, \
  and\ \bibinfo {author} {\bibfnamefont {M.}~\bibnamefont {Porrati}},\
  }\href@noop {} {\bibfield  {journal} {\bibinfo  {journal} {Physics Letters
  B}\ }\textbf {\bibinfo {volume} {485}},\ \bibinfo {pages} {208} (\bibinfo
  {year} {2000})}\BibitemShut {NoStop}%
\bibitem [{\citenamefont {Carroll}\ \emph {et~al.}(2004)\citenamefont
  {Carroll}, \citenamefont {Duvvuri}, \citenamefont {Trodden},\ and\
  \citenamefont {Turner}}]{carroll2004cosmic}%
  \BibitemOpen
  \bibfield  {author} {\bibinfo {author} {\bibfnamefont {S.~M.}\ \bibnamefont
  {Carroll}}, \bibinfo {author} {\bibfnamefont {V.}~\bibnamefont {Duvvuri}},
  \bibinfo {author} {\bibfnamefont {M.}~\bibnamefont {Trodden}}, \ and\
  \bibinfo {author} {\bibfnamefont {M.~S.}\ \bibnamefont {Turner}},\
  }\href@noop {} {\bibfield  {journal} {\bibinfo  {journal} {Physical Review
  D}\ }\textbf {\bibinfo {volume} {70}},\ \bibinfo {pages} {043528} (\bibinfo
  {year} {2004})}\BibitemShut {NoStop}%
\bibitem [{\citenamefont {Brans}\ and\ \citenamefont
  {Dicke}(1961)}]{brans1961mach}%
  \BibitemOpen
  \bibfield  {author} {\bibinfo {author} {\bibfnamefont {C.}~\bibnamefont
  {Brans}}\ and\ \bibinfo {author} {\bibfnamefont {R.~H.}\ \bibnamefont
  {Dicke}},\ }\href@noop {} {\bibfield  {journal} {\bibinfo  {journal}
  {Physical review}\ }\textbf {\bibinfo {volume} {124}},\ \bibinfo {pages}
  {925} (\bibinfo {year} {1961})}\BibitemShut {NoStop}%
\bibitem [{\citenamefont {Sahni}\ and\ \citenamefont
  {Shtanov}(2003)}]{sahni2003braneworld}%
  \BibitemOpen
  \bibfield  {author} {\bibinfo {author} {\bibfnamefont {V.}~\bibnamefont
  {Sahni}}\ and\ \bibinfo {author} {\bibfnamefont {Y.}~\bibnamefont
  {Shtanov}},\ }\href@noop {} {\bibfield  {journal} {\bibinfo  {journal}
  {Journal of Cosmology and Astroparticle Physics}\ }\textbf {\bibinfo {volume}
  {2003}},\ \bibinfo {pages} {014} (\bibinfo {year} {2003})}\BibitemShut
  {NoStop}%
\bibitem [{\citenamefont {Aghanim}\ \emph {et~al.}(2018)\citenamefont
  {Aghanim}, \citenamefont {Akrami}, \citenamefont {Ashdown}, \citenamefont
  {Aumont}, \citenamefont {Baccigalupi}, \citenamefont {Ballardini},
  \citenamefont {Banday}, \citenamefont {Barreiro}, \citenamefont {Bartolo},
  \citenamefont {Basak} \emph {et~al.}}]{aghanim2018planck}%
  \BibitemOpen
  \bibfield  {author} {\bibinfo {author} {\bibfnamefont {N.}~\bibnamefont
  {Aghanim}}, \bibinfo {author} {\bibfnamefont {Y.}~\bibnamefont {Akrami}},
  \bibinfo {author} {\bibfnamefont {M.}~\bibnamefont {Ashdown}}, \bibinfo
  {author} {\bibfnamefont {J.}~\bibnamefont {Aumont}}, \bibinfo {author}
  {\bibfnamefont {C.}~\bibnamefont {Baccigalupi}}, \bibinfo {author}
  {\bibfnamefont {M.}~\bibnamefont {Ballardini}}, \bibinfo {author}
  {\bibfnamefont {A.}~\bibnamefont {Banday}}, \bibinfo {author} {\bibfnamefont
  {R.}~\bibnamefont {Barreiro}}, \bibinfo {author} {\bibfnamefont
  {N.}~\bibnamefont {Bartolo}}, \bibinfo {author} {\bibfnamefont
  {S.}~\bibnamefont {Basak}},  \emph {et~al.},\ }\href@noop {} {\bibfield
  {journal} {\bibinfo  {journal} {arXiv preprint arXiv:1807.06209}\ } (\bibinfo
  {year} {2018})}\BibitemShut {NoStop}%
\bibitem [{\citenamefont {Carroll}(2001)}]{carroll2001cosmological}%
  \BibitemOpen
  \bibfield  {author} {\bibinfo {author} {\bibfnamefont {S.~M.}\ \bibnamefont
  {Carroll}},\ }\href@noop {} {\bibfield  {journal} {\bibinfo  {journal}
  {Living Reviews in Relativity}\ }\textbf {\bibinfo {volume} {4}},\ \bibinfo
  {pages} {1} (\bibinfo {year} {2001})}\BibitemShut {NoStop}%
\bibitem [{\citenamefont {Weinberg}(1989)}]{weinberg1989cosmological}%
  \BibitemOpen
  \bibfield  {author} {\bibinfo {author} {\bibfnamefont {S.}~\bibnamefont
  {Weinberg}},\ }\href@noop {} {\bibfield  {journal} {\bibinfo  {journal}
  {Reviews of modern physics}\ }\textbf {\bibinfo {volume} {61}},\ \bibinfo
  {pages} {1} (\bibinfo {year} {1989})}\BibitemShut {NoStop}%
\bibitem [{\citenamefont {Riess}\ \emph {et~al.}(2019)\citenamefont {Riess},
  \citenamefont {Casertano}, \citenamefont {Yuan}, \citenamefont {Macri},\ and\
  \citenamefont {Scolnic}}]{riess2019Large}%
  \BibitemOpen
  \bibfield  {author} {\bibinfo {author} {\bibfnamefont {A.~G.}\ \bibnamefont
  {Riess}}, \bibinfo {author} {\bibfnamefont {S.}~\bibnamefont {Casertano}},
  \bibinfo {author} {\bibfnamefont {W.}~\bibnamefont {Yuan}}, \bibinfo {author}
  {\bibfnamefont {L.~M.}\ \bibnamefont {Macri}}, \ and\ \bibinfo {author}
  {\bibfnamefont {D.}~\bibnamefont {Scolnic}},\ }\href@noop {} {\bibfield
  {journal} {\bibinfo  {journal} {arXiv preprint arXiv:1903.07603}\ } (\bibinfo
  {year} {2019})}\BibitemShut {NoStop}%
\bibitem [{\citenamefont {Caldwell}\ \emph {et~al.}(1998)\citenamefont
  {Caldwell}, \citenamefont {Dave},\ and\ \citenamefont
  {Steinhardt}}]{1998cosmological}%
  \BibitemOpen
  \bibfield  {author} {\bibinfo {author} {\bibfnamefont {R.~R.}\ \bibnamefont
  {Caldwell}}, \bibinfo {author} {\bibfnamefont {R.}~\bibnamefont {Dave}}, \
  and\ \bibinfo {author} {\bibfnamefont {P.~J.}\ \bibnamefont {Steinhardt}},\
  }\href@noop {} {\bibfield  {journal} {\bibinfo  {journal} {Physical Review
  Letters}\ }\textbf {\bibinfo {volume} {80}},\ \bibinfo {pages} {1582}
  (\bibinfo {year} {1998})}\BibitemShut {NoStop}%
\bibitem [{\citenamefont {Kamenshchik}\ \emph {et~al.}(2001)\citenamefont
  {Kamenshchik}, \citenamefont {Moschella},\ and\ \citenamefont
  {Pasquier}}]{kamenshchik2001alternative}%
  \BibitemOpen
  \bibfield  {author} {\bibinfo {author} {\bibfnamefont {A.}~\bibnamefont
  {Kamenshchik}}, \bibinfo {author} {\bibfnamefont {U.}~\bibnamefont
  {Moschella}}, \ and\ \bibinfo {author} {\bibfnamefont {V.}~\bibnamefont
  {Pasquier}},\ }\href@noop {} {\bibfield  {journal} {\bibinfo  {journal}
  {Physics Letters B}\ }\textbf {\bibinfo {volume} {511}},\ \bibinfo {pages}
  {265} (\bibinfo {year} {2001})}\BibitemShut {NoStop}%
\bibitem [{\citenamefont {Amendola}(2000)}]{amendola2000coupled}%
  \BibitemOpen
  \bibfield  {author} {\bibinfo {author} {\bibfnamefont {L.}~\bibnamefont
  {Amendola}},\ }\href@noop {} {\bibfield  {journal} {\bibinfo  {journal}
  {Physical Review D}\ }\textbf {\bibinfo {volume} {62}},\ \bibinfo {pages}
  {043511} (\bibinfo {year} {2000})}\BibitemShut {NoStop}%
\bibitem [{\citenamefont {Chiba}\ \emph {et~al.}(2000)\citenamefont {Chiba},
  \citenamefont {Okabe},\ and\ \citenamefont
  {Yamaguchi}}]{chiba2000kinetically}%
  \BibitemOpen
  \bibfield  {author} {\bibinfo {author} {\bibfnamefont {T.}~\bibnamefont
  {Chiba}}, \bibinfo {author} {\bibfnamefont {T.}~\bibnamefont {Okabe}}, \ and\
  \bibinfo {author} {\bibfnamefont {M.}~\bibnamefont {Yamaguchi}},\ }\href@noop
  {} {\bibfield  {journal} {\bibinfo  {journal} {Physical Review D}\ }\textbf
  {\bibinfo {volume} {62}},\ \bibinfo {pages} {023511} (\bibinfo {year}
  {2000})}\BibitemShut {NoStop}%
\bibitem [{\citenamefont {Zlatev}\ \emph {et~al.}(1999)\citenamefont {Zlatev},
  \citenamefont {Wang},\ and\ \citenamefont
  {Steinhardt}}]{zlatev1999quintessence}%
  \BibitemOpen
  \bibfield  {author} {\bibinfo {author} {\bibfnamefont {I.}~\bibnamefont
  {Zlatev}}, \bibinfo {author} {\bibfnamefont {L.}~\bibnamefont {Wang}}, \ and\
  \bibinfo {author} {\bibfnamefont {P.~J.}\ \bibnamefont {Steinhardt}},\
  }\href@noop {} {\bibfield  {journal} {\bibinfo  {journal} {Physical Review
  Letters}\ }\textbf {\bibinfo {volume} {82}},\ \bibinfo {pages} {896}
  (\bibinfo {year} {1999})}\BibitemShut {NoStop}%
\bibitem [{\citenamefont {Caldwell}(2002)}]{caldwell2002phantom}%
  \BibitemOpen
  \bibfield  {author} {\bibinfo {author} {\bibfnamefont {R.~R.}\ \bibnamefont
  {Caldwell}},\ }\href@noop {} {\bibfield  {journal} {\bibinfo  {journal}
  {Physics Letters B}\ }\textbf {\bibinfo {volume} {545}},\ \bibinfo {pages}
  {23} (\bibinfo {year} {2002})}\BibitemShut {NoStop}%
\bibitem [{\citenamefont {Caldwell}\ \emph {et~al.}(2003)\citenamefont
  {Caldwell}, \citenamefont {Kamionkowski},\ and\ \citenamefont
  {Weinberg}}]{2003phantom}%
  \BibitemOpen
  \bibfield  {author} {\bibinfo {author} {\bibfnamefont {R.~R.}\ \bibnamefont
  {Caldwell}}, \bibinfo {author} {\bibfnamefont {M.}~\bibnamefont
  {Kamionkowski}}, \ and\ \bibinfo {author} {\bibfnamefont {N.~N.}\
  \bibnamefont {Weinberg}},\ }\href@noop {} {\bibfield  {journal} {\bibinfo
  {journal} {Physical Review Letters}\ }\textbf {\bibinfo {volume} {91}},\
  \bibinfo {pages} {071301} (\bibinfo {year} {2003})}\BibitemShut {NoStop}%
\bibitem [{\citenamefont {Nojiri}\ and\ \citenamefont
  {Odintsov}(2003)}]{nojiri2003quantum}%
  \BibitemOpen
  \bibfield  {author} {\bibinfo {author} {\bibfnamefont {S.}~\bibnamefont
  {Nojiri}}\ and\ \bibinfo {author} {\bibfnamefont {S.~D.}\ \bibnamefont
  {Odintsov}},\ }\href@noop {} {\bibfield  {journal} {\bibinfo  {journal}
  {Physics Letters B}\ }\textbf {\bibinfo {volume} {562}},\ \bibinfo {pages}
  {147} (\bibinfo {year} {2003})}\BibitemShut {NoStop}%
\bibitem [{\citenamefont {Armendariz-Picon}\ \emph {et~al.}(2001)\citenamefont
  {Armendariz-Picon}, \citenamefont {Mukhanov},\ and\ \citenamefont
  {Steinhardt}}]{armendariz2001essentials}%
  \BibitemOpen
  \bibfield  {author} {\bibinfo {author} {\bibfnamefont {C.}~\bibnamefont
  {Armendariz-Picon}}, \bibinfo {author} {\bibfnamefont {V.}~\bibnamefont
  {Mukhanov}}, \ and\ \bibinfo {author} {\bibfnamefont {P.~J.}\ \bibnamefont
  {Steinhardt}},\ }\href@noop {} {\bibfield  {journal} {\bibinfo  {journal}
  {Physical Review D}\ }\textbf {\bibinfo {volume} {63}},\ \bibinfo {pages}
  {103510} (\bibinfo {year} {2001})}\BibitemShut {NoStop}%
\bibitem [{\citenamefont {Deffayet}\ \emph {et~al.}(2011)\citenamefont
  {Deffayet}, \citenamefont {Gao}, \citenamefont {Steer},\ and\ \citenamefont
  {Zahariade}}]{deffayet2011k}%
  \BibitemOpen
  \bibfield  {author} {\bibinfo {author} {\bibfnamefont {C.}~\bibnamefont
  {Deffayet}}, \bibinfo {author} {\bibfnamefont {X.}~\bibnamefont {Gao}},
  \bibinfo {author} {\bibfnamefont {D.~A.}\ \bibnamefont {Steer}}, \ and\
  \bibinfo {author} {\bibfnamefont {G.}~\bibnamefont {Zahariade}},\ }\href@noop
  {} {\bibfield  {journal} {\bibinfo  {journal} {Physical Review D}\ }\textbf
  {\bibinfo {volume} {84}},\ \bibinfo {pages} {064039} (\bibinfo {year}
  {2011})}\BibitemShut {NoStop}%
\bibitem [{\citenamefont {Scherrer}(2004)}]{scherrer2004purely}%
  \BibitemOpen
  \bibfield  {author} {\bibinfo {author} {\bibfnamefont {R.~J.}\ \bibnamefont
  {Scherrer}},\ }\href@noop {} {\bibfield  {journal} {\bibinfo  {journal}
  {Physical review letters}\ }\textbf {\bibinfo {volume} {93}},\ \bibinfo
  {pages} {011301} (\bibinfo {year} {2004})}\BibitemShut {NoStop}%
\bibitem [{\citenamefont {Cai}\ \emph {et~al.}(2010)\citenamefont {Cai},
  \citenamefont {Saridakis}, \citenamefont {Setare},\ and\ \citenamefont
  {Xia}}]{cai2010quintom}%
  \BibitemOpen
  \bibfield  {author} {\bibinfo {author} {\bibfnamefont {Y.-F.}\ \bibnamefont
  {Cai}}, \bibinfo {author} {\bibfnamefont {E.~N.}\ \bibnamefont {Saridakis}},
  \bibinfo {author} {\bibfnamefont {M.~R.}\ \bibnamefont {Setare}}, \ and\
  \bibinfo {author} {\bibfnamefont {J.-Q.}\ \bibnamefont {Xia}},\ }\href@noop
  {} {\bibfield  {journal} {\bibinfo  {journal} {Physics Reports}\ }\textbf
  {\bibinfo {volume} {493}},\ \bibinfo {pages} {1} (\bibinfo {year}
  {2010})}\BibitemShut {NoStop}%
\bibitem [{\citenamefont {Guo}\ \emph {et~al.}(2005)\citenamefont {Guo},
  \citenamefont {Piao}, \citenamefont {Zhang},\ and\ \citenamefont
  {Zhang}}]{guo2005cosmological}%
  \BibitemOpen
  \bibfield  {author} {\bibinfo {author} {\bibfnamefont {Z.-K.}\ \bibnamefont
  {Guo}}, \bibinfo {author} {\bibfnamefont {Y.-S.}\ \bibnamefont {Piao}},
  \bibinfo {author} {\bibfnamefont {X.}~\bibnamefont {Zhang}}, \ and\ \bibinfo
  {author} {\bibfnamefont {Y.-Z.}\ \bibnamefont {Zhang}},\ }\href@noop {}
  {\bibfield  {journal} {\bibinfo  {journal} {Physics Letters B}\ }\textbf
  {\bibinfo {volume} {608}},\ \bibinfo {pages} {177} (\bibinfo {year}
  {2005})}\BibitemShut {NoStop}%
\bibitem [{\citenamefont {Bagla}\ \emph {et~al.}(2003)\citenamefont {Bagla},
  \citenamefont {Jassal},\ and\ \citenamefont
  {Padmanabhan}}]{bagla2003cosmology}%
  \BibitemOpen
  \bibfield  {author} {\bibinfo {author} {\bibfnamefont {J.}~\bibnamefont
  {Bagla}}, \bibinfo {author} {\bibfnamefont {H.~K.}\ \bibnamefont {Jassal}}, \
  and\ \bibinfo {author} {\bibfnamefont {T.}~\bibnamefont {Padmanabhan}},\
  }\href@noop {} {\bibfield  {journal} {\bibinfo  {journal} {Physical Review
  D}\ }\textbf {\bibinfo {volume} {67}},\ \bibinfo {pages} {063504} (\bibinfo
  {year} {2003})}\BibitemShut {NoStop}%
\bibitem [{\citenamefont {Li}(2004)}]{li2004model}%
  \BibitemOpen
  \bibfield  {author} {\bibinfo {author} {\bibfnamefont {M.}~\bibnamefont
  {Li}},\ }\href@noop {} {\bibfield  {journal} {\bibinfo  {journal} {Physics
  Letters B}\ }\textbf {\bibinfo {volume} {603}},\ \bibinfo {pages} {1}
  (\bibinfo {year} {2004})}\BibitemShut {NoStop}%
\bibitem [{\citenamefont {Wei}\ and\ \citenamefont {Cai}(2008)}]{wei2008new}%
  \BibitemOpen
  \bibfield  {author} {\bibinfo {author} {\bibfnamefont {H.}~\bibnamefont
  {Wei}}\ and\ \bibinfo {author} {\bibfnamefont {R.-G.}\ \bibnamefont {Cai}},\
  }\href@noop {} {\bibfield  {journal} {\bibinfo  {journal} {Physics Letters
  B}\ }\textbf {\bibinfo {volume} {660}},\ \bibinfo {pages} {113} (\bibinfo
  {year} {2008})}\BibitemShut {NoStop}%
\bibitem [{\citenamefont {Bento}\ \emph {et~al.}(2002)\citenamefont {Bento},
  \citenamefont {Bertolami},\ and\ \citenamefont {Sen}}]{bento2002generalized}%
  \BibitemOpen
  \bibfield  {author} {\bibinfo {author} {\bibfnamefont {M.}~\bibnamefont
  {Bento}}, \bibinfo {author} {\bibfnamefont {O.}~\bibnamefont {Bertolami}}, \
  and\ \bibinfo {author} {\bibfnamefont {A.~A.}\ \bibnamefont {Sen}},\
  }\href@noop {} {\bibfield  {journal} {\bibinfo  {journal} {Physical Review
  D}\ }\textbf {\bibinfo {volume} {66}},\ \bibinfo {pages} {043507} (\bibinfo
  {year} {2002})}\BibitemShut {NoStop}%
\bibitem [{\citenamefont {Gorini}\ \emph {et~al.}(2003)\citenamefont {Gorini},
  \citenamefont {Kamenshchik},\ and\ \citenamefont
  {Moschella}}]{gorini2003can}%
  \BibitemOpen
  \bibfield  {author} {\bibinfo {author} {\bibfnamefont {V.}~\bibnamefont
  {Gorini}}, \bibinfo {author} {\bibfnamefont {A.}~\bibnamefont {Kamenshchik}},
  \ and\ \bibinfo {author} {\bibfnamefont {U.}~\bibnamefont {Moschella}},\
  }\href@noop {} {\bibfield  {journal} {\bibinfo  {journal} {Physical Review
  D}\ }\textbf {\bibinfo {volume} {67}},\ \bibinfo {pages} {063509} (\bibinfo
  {year} {2003})}\BibitemShut {NoStop}%
\bibitem [{\citenamefont {Chevallier}\ and\ \citenamefont
  {Polarski}(2001)}]{chevallier2001accelerating}%
  \BibitemOpen
  \bibfield  {author} {\bibinfo {author} {\bibfnamefont {M.}~\bibnamefont
  {Chevallier}}\ and\ \bibinfo {author} {\bibfnamefont {D.}~\bibnamefont
  {Polarski}},\ }\href@noop {} {\bibfield  {journal} {\bibinfo  {journal}
  {International Journal of Modern Physics D}\ }\textbf {\bibinfo {volume}
  {10}},\ \bibinfo {pages} {213} (\bibinfo {year} {2001})}\BibitemShut
  {NoStop}%
\bibitem [{\citenamefont {Linder}(2003)}]{linder2003exploring}%
  \BibitemOpen
  \bibfield  {author} {\bibinfo {author} {\bibfnamefont {E.~V.}\ \bibnamefont
  {Linder}},\ }\href@noop {} {\bibfield  {journal} {\bibinfo  {journal}
  {Physical Review Letters}\ }\textbf {\bibinfo {volume} {90}},\ \bibinfo
  {pages} {091301} (\bibinfo {year} {2003})}\BibitemShut {NoStop}%
\bibitem [{\citenamefont {Barboza~Jr}\ \emph {et~al.}(2009)\citenamefont
  {Barboza~Jr}, \citenamefont {Alcaniz}, \citenamefont {Zhu},\ and\
  \citenamefont {Silva}}]{barboza2009generalized}%
  \BibitemOpen
  \bibfield  {author} {\bibinfo {author} {\bibfnamefont {E.}~\bibnamefont
  {Barboza~Jr}}, \bibinfo {author} {\bibfnamefont {J.}~\bibnamefont {Alcaniz}},
  \bibinfo {author} {\bibfnamefont {Z.-H.}\ \bibnamefont {Zhu}}, \ and\
  \bibinfo {author} {\bibfnamefont {R.}~\bibnamefont {Silva}},\ }\href@noop {}
  {\bibfield  {journal} {\bibinfo  {journal} {Physical Review D}\ }\textbf
  {\bibinfo {volume} {80}},\ \bibinfo {pages} {043521} (\bibinfo {year}
  {2009})}\BibitemShut {NoStop}%
\bibitem [{\citenamefont {Catto{\"e}n}\ and\ \citenamefont
  {Visser}(2007)}]{cattoen2007hubble}%
  \BibitemOpen
  \bibfield  {author} {\bibinfo {author} {\bibfnamefont {C.}~\bibnamefont
  {Catto{\"e}n}}\ and\ \bibinfo {author} {\bibfnamefont {M.}~\bibnamefont
  {Visser}},\ }\href@noop {} {\bibfield  {journal} {\bibinfo  {journal}
  {Classical and Quantum Gravity}\ }\textbf {\bibinfo {volume} {24}},\ \bibinfo
  {pages} {5985} (\bibinfo {year} {2007})}\BibitemShut {NoStop}%
\bibitem [{\citenamefont {Gruber}\ and\ \citenamefont
  {Luongo}(2014)}]{gruber2014cosmographic}%
  \BibitemOpen
  \bibfield  {author} {\bibinfo {author} {\bibfnamefont {C.}~\bibnamefont
  {Gruber}}\ and\ \bibinfo {author} {\bibfnamefont {O.}~\bibnamefont
  {Luongo}},\ }\href@noop {} {\bibfield  {journal} {\bibinfo  {journal}
  {Physical Review D}\ }\textbf {\bibinfo {volume} {89}},\ \bibinfo {pages}
  {103506} (\bibinfo {year} {2014})}\BibitemShut {NoStop}%
\bibitem [{\citenamefont {Alam}\ \emph {et~al.}(2004)\citenamefont {Alam},
  \citenamefont {Sahni}, \citenamefont {Deep~Saini},\ and\ \citenamefont
  {Starobinsky}}]{alam2004there}%
  \BibitemOpen
  \bibfield  {author} {\bibinfo {author} {\bibfnamefont {U.}~\bibnamefont
  {Alam}}, \bibinfo {author} {\bibfnamefont {V.}~\bibnamefont {Sahni}},
  \bibinfo {author} {\bibfnamefont {T.}~\bibnamefont {Deep~Saini}}, \ and\
  \bibinfo {author} {\bibfnamefont {A.~A.}\ \bibnamefont {Starobinsky}},\
  }\href@noop {} {\bibfield  {journal} {\bibinfo  {journal} {Monthly Notices of
  the Royal Astronomical Society}\ }\textbf {\bibinfo {volume} {354}},\
  \bibinfo {pages} {275} (\bibinfo {year} {2004})}\BibitemShut {NoStop}%
\bibitem [{\citenamefont {Sen}(2008)}]{sen2008deviation}%
  \BibitemOpen
  \bibfield  {author} {\bibinfo {author} {\bibfnamefont {A.~A.}\ \bibnamefont
  {Sen}},\ }\href@noop {} {\bibfield  {journal} {\bibinfo  {journal} {Physical
  Review D}\ }\textbf {\bibinfo {volume} {77}},\ \bibinfo {pages} {043508}
  (\bibinfo {year} {2008})}\BibitemShut {NoStop}%
\bibitem [{\citenamefont {Kumar}\ \emph {et~al.}(2013)\citenamefont {Kumar},
  \citenamefont {Nautiyal},\ and\ \citenamefont {Sen}}]{kumar2013deviation}%
  \BibitemOpen
  \bibfield  {author} {\bibinfo {author} {\bibfnamefont {S.}~\bibnamefont
  {Kumar}}, \bibinfo {author} {\bibfnamefont {A.}~\bibnamefont {Nautiyal}}, \
  and\ \bibinfo {author} {\bibfnamefont {A.~A.}\ \bibnamefont {Sen}},\
  }\href@noop {} {\bibfield  {journal} {\bibinfo  {journal} {The European
  Physical Journal C}\ }\textbf {\bibinfo {volume} {73}},\ \bibinfo {pages}
  {2562} (\bibinfo {year} {2013})}\BibitemShut {NoStop}%
\bibitem [{\citenamefont {Zhang}\ \emph {et~al.}(2015)\citenamefont {Zhang},
  \citenamefont {Yang}, \citenamefont {Zou}, \citenamefont {Meng},\ and\
  \citenamefont {Shen}}]{zhang2015exploring}%
  \BibitemOpen
  \bibfield  {author} {\bibinfo {author} {\bibfnamefont {Q.}~\bibnamefont
  {Zhang}}, \bibinfo {author} {\bibfnamefont {G.}~\bibnamefont {Yang}},
  \bibinfo {author} {\bibfnamefont {Q.}~\bibnamefont {Zou}}, \bibinfo {author}
  {\bibfnamefont {X.}~\bibnamefont {Meng}}, \ and\ \bibinfo {author}
  {\bibfnamefont {K.}~\bibnamefont {Shen}},\ }\href@noop {} {\bibfield
  {journal} {\bibinfo  {journal} {The European Physical Journal C}\ }\textbf
  {\bibinfo {volume} {75}},\ \bibinfo {pages} {300} (\bibinfo {year}
  {2015})}\BibitemShut {NoStop}%
\bibitem [{\citenamefont {Yang}\ \emph {et~al.}(2016)\citenamefont {Yang},
  \citenamefont {Wang},\ and\ \citenamefont {Meng}}]{yang2016diagnostics}%
  \BibitemOpen
  \bibfield  {author} {\bibinfo {author} {\bibfnamefont {G.}~\bibnamefont
  {Yang}}, \bibinfo {author} {\bibfnamefont {D.}~\bibnamefont {Wang}}, \ and\
  \bibinfo {author} {\bibfnamefont {X.}~\bibnamefont {Meng}},\ }\href@noop {}
  {\bibfield  {journal} {\bibinfo  {journal} {arXiv preprint arXiv:1602.02552}\
  } (\bibinfo {year} {2016})}\BibitemShut {NoStop}%
\bibitem [{\citenamefont {Wang}\ \emph
  {et~al.}(2017{\natexlab{a}})\citenamefont {Wang}, \citenamefont {Yan},\ and\
  \citenamefont {Meng}}]{wang2017new}%
  \BibitemOpen
  \bibfield  {author} {\bibinfo {author} {\bibfnamefont {D.}~\bibnamefont
  {Wang}}, \bibinfo {author} {\bibfnamefont {Y.-J.}\ \bibnamefont {Yan}}, \
  and\ \bibinfo {author} {\bibfnamefont {X.-H.}\ \bibnamefont {Meng}},\
  }\href@noop {} {\bibfield  {journal} {\bibinfo  {journal} {The European
  Physical Journal C}\ }\textbf {\bibinfo {volume} {77}},\ \bibinfo {pages}
  {263} (\bibinfo {year} {2017}{\natexlab{a}})}\BibitemShut {NoStop}%
\bibitem [{\citenamefont {Wang}\ and\ \citenamefont
  {Meng}(2018)}]{wang2018pressure}%
  \BibitemOpen
  \bibfield  {author} {\bibinfo {author} {\bibfnamefont {J.-C.}\ \bibnamefont
  {Wang}}\ and\ \bibinfo {author} {\bibfnamefont {X.-H.}\ \bibnamefont
  {Meng}},\ }\href@noop {} {\bibfield  {journal} {\bibinfo  {journal}
  {Communications in Theoretical Physics}\ }\textbf {\bibinfo {volume} {70}},\
  \bibinfo {pages} {713} (\bibinfo {year} {2018})}\BibitemShut {NoStop}%
\bibitem [{\citenamefont {Akarsu}\ and\ \citenamefont
  {Dereli}(2012)}]{akarsu2012cosmological}%
  \BibitemOpen
  \bibfield  {author} {\bibinfo {author} {\bibfnamefont {{\"O}.}~\bibnamefont
  {Akarsu}}\ and\ \bibinfo {author} {\bibfnamefont {T.}~\bibnamefont
  {Dereli}},\ }\href@noop {} {\bibfield  {journal} {\bibinfo  {journal}
  {International Journal of Theoretical Physics}\ }\textbf {\bibinfo {volume}
  {51}},\ \bibinfo {pages} {612} (\bibinfo {year} {2012})}\BibitemShut
  {NoStop}%
\bibitem [{\citenamefont {Zhang}\ \emph {et~al.}(2014)\citenamefont {Zhang},
  \citenamefont {Zhang}, \citenamefont {Yuan}, \citenamefont {Liu},
  \citenamefont {Zhang},\ and\ \citenamefont {Sun}}]{zhang2014four}%
  \BibitemOpen
  \bibfield  {author} {\bibinfo {author} {\bibfnamefont {C.}~\bibnamefont
  {Zhang}}, \bibinfo {author} {\bibfnamefont {H.}~\bibnamefont {Zhang}},
  \bibinfo {author} {\bibfnamefont {S.}~\bibnamefont {Yuan}}, \bibinfo {author}
  {\bibfnamefont {S.}~\bibnamefont {Liu}}, \bibinfo {author} {\bibfnamefont
  {T.-J.}\ \bibnamefont {Zhang}}, \ and\ \bibinfo {author} {\bibfnamefont
  {Y.-C.}\ \bibnamefont {Sun}},\ }\href@noop {} {\bibfield  {journal} {\bibinfo
   {journal} {Research in Astronomy and Astrophysics}\ }\textbf {\bibinfo
  {volume} {14}},\ \bibinfo {pages} {1221} (\bibinfo {year}
  {2014})}\BibitemShut {NoStop}%
\bibitem [{\citenamefont {Moresco}\ \emph {et~al.}(2012)\citenamefont
  {Moresco}, \citenamefont {Cimatti}, \citenamefont {Jimenez}, \citenamefont
  {Pozzetti}, \citenamefont {Zamorani}, \citenamefont {Bolzonella},
  \citenamefont {Dunlop}, \citenamefont {Lamareille}, \citenamefont {Mignoli},
  \citenamefont {Pearce} \emph {et~al.}}]{moresco2012improved}%
  \BibitemOpen
  \bibfield  {author} {\bibinfo {author} {\bibfnamefont {M.}~\bibnamefont
  {Moresco}}, \bibinfo {author} {\bibfnamefont {A.}~\bibnamefont {Cimatti}},
  \bibinfo {author} {\bibfnamefont {R.}~\bibnamefont {Jimenez}}, \bibinfo
  {author} {\bibfnamefont {L.}~\bibnamefont {Pozzetti}}, \bibinfo {author}
  {\bibfnamefont {G.}~\bibnamefont {Zamorani}}, \bibinfo {author}
  {\bibfnamefont {M.}~\bibnamefont {Bolzonella}}, \bibinfo {author}
  {\bibfnamefont {J.}~\bibnamefont {Dunlop}}, \bibinfo {author} {\bibfnamefont
  {F.}~\bibnamefont {Lamareille}}, \bibinfo {author} {\bibfnamefont
  {M.}~\bibnamefont {Mignoli}}, \bibinfo {author} {\bibfnamefont
  {H.}~\bibnamefont {Pearce}},  \emph {et~al.},\ }\href@noop {} {\bibfield
  {journal} {\bibinfo  {journal} {Journal of Cosmology and Astroparticle
  Physics}\ }\textbf {\bibinfo {volume} {2012}},\ \bibinfo {pages} {006}
  (\bibinfo {year} {2012})}\BibitemShut {NoStop}%
\bibitem [{\citenamefont {Jimenez}\ \emph {et~al.}(2003)\citenamefont
  {Jimenez}, \citenamefont {Verde}, \citenamefont {Treu},\ and\ \citenamefont
  {Stern}}]{jimenez2003constraints}%
  \BibitemOpen
  \bibfield  {author} {\bibinfo {author} {\bibfnamefont {R.}~\bibnamefont
  {Jimenez}}, \bibinfo {author} {\bibfnamefont {L.}~\bibnamefont {Verde}},
  \bibinfo {author} {\bibfnamefont {T.}~\bibnamefont {Treu}}, \ and\ \bibinfo
  {author} {\bibfnamefont {D.}~\bibnamefont {Stern}},\ }\href@noop {}
  {\bibfield  {journal} {\bibinfo  {journal} {The Astrophysical Journal}\
  }\textbf {\bibinfo {volume} {593}},\ \bibinfo {pages} {622} (\bibinfo {year}
  {2003})}\BibitemShut {NoStop}%
\bibitem [{\citenamefont {Moresco}\ \emph {et~al.}(2016)\citenamefont
  {Moresco}, \citenamefont {Pozzetti}, \citenamefont {Cimatti}, \citenamefont
  {Jimenez}, \citenamefont {Maraston}, \citenamefont {Verde}, \citenamefont
  {Thomas}, \citenamefont {Citro}, \citenamefont {Tojeiro},\ and\ \citenamefont
  {Wilkinson}}]{moresco20166}%
  \BibitemOpen
  \bibfield  {author} {\bibinfo {author} {\bibfnamefont {M.}~\bibnamefont
  {Moresco}}, \bibinfo {author} {\bibfnamefont {L.}~\bibnamefont {Pozzetti}},
  \bibinfo {author} {\bibfnamefont {A.}~\bibnamefont {Cimatti}}, \bibinfo
  {author} {\bibfnamefont {R.}~\bibnamefont {Jimenez}}, \bibinfo {author}
  {\bibfnamefont {C.}~\bibnamefont {Maraston}}, \bibinfo {author}
  {\bibfnamefont {L.}~\bibnamefont {Verde}}, \bibinfo {author} {\bibfnamefont
  {D.}~\bibnamefont {Thomas}}, \bibinfo {author} {\bibfnamefont
  {A.}~\bibnamefont {Citro}}, \bibinfo {author} {\bibfnamefont
  {R.}~\bibnamefont {Tojeiro}}, \ and\ \bibinfo {author} {\bibfnamefont
  {D.}~\bibnamefont {Wilkinson}},\ }\href@noop {} {\bibfield  {journal}
  {\bibinfo  {journal} {Journal of Cosmology and Astroparticle Physics}\
  }\textbf {\bibinfo {volume} {2016}},\ \bibinfo {pages} {014} (\bibinfo {year}
  {2016})}\BibitemShut {NoStop}%
\bibitem [{\citenamefont {Stern}()}]{stern2010d}%
  \BibitemOpen
  \bibfield  {author} {\bibinfo {author} {\bibfnamefont {D.}~\bibnamefont
  {Stern}},\ }\href@noop {} {\bibfield  {journal} {\bibinfo  {journal} {J.
  Cosmol. Astropart. Phys.}\ }\textbf {\bibinfo {volume} {2010}},\ \bibinfo
  {pages} {008}}\BibitemShut {NoStop}%
\bibitem [{\citenamefont {Simon}\ \emph {et~al.}(2005)\citenamefont {Simon},
  \citenamefont {Verde},\ and\ \citenamefont {Jimenez}}]{simon2005constraints}%
  \BibitemOpen
  \bibfield  {author} {\bibinfo {author} {\bibfnamefont {J.}~\bibnamefont
  {Simon}}, \bibinfo {author} {\bibfnamefont {L.}~\bibnamefont {Verde}}, \ and\
  \bibinfo {author} {\bibfnamefont {R.}~\bibnamefont {Jimenez}},\ }\href@noop
  {} {\bibfield  {journal} {\bibinfo  {journal} {Physical Review D}\ }\textbf
  {\bibinfo {volume} {71}},\ \bibinfo {pages} {123001} (\bibinfo {year}
  {2005})}\BibitemShut {NoStop}%
\bibitem [{\citenamefont {Wang}\ \emph
  {et~al.}(2017{\natexlab{b}})\citenamefont {Wang}, \citenamefont {Zhao},
  \citenamefont {Chuang}, \citenamefont {Pellejero-Ibanez}, \citenamefont
  {Zhao}, \citenamefont {Kitaura},\ and\ \citenamefont
  {Rodriguez-Torres}}]{wang2017clustering}%
  \BibitemOpen
  \bibfield  {author} {\bibinfo {author} {\bibfnamefont {Y.}~\bibnamefont
  {Wang}}, \bibinfo {author} {\bibfnamefont {G.-B.}\ \bibnamefont {Zhao}},
  \bibinfo {author} {\bibfnamefont {C.-H.}\ \bibnamefont {Chuang}}, \bibinfo
  {author} {\bibfnamefont {M.}~\bibnamefont {Pellejero-Ibanez}}, \bibinfo
  {author} {\bibfnamefont {C.}~\bibnamefont {Zhao}}, \bibinfo {author}
  {\bibfnamefont {F.-S.}\ \bibnamefont {Kitaura}}, \ and\ \bibinfo {author}
  {\bibfnamefont {S.}~\bibnamefont {Rodriguez-Torres}},\ }\href@noop {}
  {\bibfield  {journal} {\bibinfo  {journal} {arXiv preprint arXiv:1709.05173}\
  } (\bibinfo {year} {2017}{\natexlab{b}})}\BibitemShut {NoStop}%
\bibitem [{\citenamefont {Moresco}(2015)}]{moresco2015raising}%
  \BibitemOpen
  \bibfield  {author} {\bibinfo {author} {\bibfnamefont {M.}~\bibnamefont
  {Moresco}},\ }\href@noop {} {\bibfield  {journal} {\bibinfo  {journal}
  {Monthly Notices of the Royal Astronomical Society: Letters}\ }\textbf
  {\bibinfo {volume} {450}},\ \bibinfo {pages} {L16} (\bibinfo {year}
  {2015})}\BibitemShut {NoStop}%
\bibitem [{\citenamefont {Foreman-Mackey}\ \emph {et~al.}(2013)\citenamefont
  {Foreman-Mackey}, \citenamefont {Hogg}, \citenamefont {Lang},\ and\
  \citenamefont {Goodman}}]{foreman2013emcee}%
  \BibitemOpen
  \bibfield  {author} {\bibinfo {author} {\bibfnamefont {D.}~\bibnamefont
  {Foreman-Mackey}}, \bibinfo {author} {\bibfnamefont {D.~W.}\ \bibnamefont
  {Hogg}}, \bibinfo {author} {\bibfnamefont {D.}~\bibnamefont {Lang}}, \ and\
  \bibinfo {author} {\bibfnamefont {J.}~\bibnamefont {Goodman}},\ }\href@noop
  {} {\bibfield  {journal} {\bibinfo  {journal} {Publications of the
  Astronomical Society of the Pacific}\ }\textbf {\bibinfo {volume} {125}},\
  \bibinfo {pages} {306} (\bibinfo {year} {2013})}\BibitemShut {NoStop}%
\bibitem [{\citenamefont {Bocquet}\ and\ \citenamefont
  {Carter}(2016)}]{Bocquet2016}%
  \BibitemOpen
  \bibfield  {author} {\bibinfo {author} {\bibfnamefont {S.}~\bibnamefont
  {Bocquet}}\ and\ \bibinfo {author} {\bibfnamefont {F.~W.}\ \bibnamefont
  {Carter}},\ }\href {\doibase 10.21105/joss.00046} {\bibfield  {journal}
  {\bibinfo  {journal} {The Journal of Open Source Software}\ }\textbf
  {\bibinfo {volume} {1}} (\bibinfo {year} {2016}),\
  10.21105/joss.00046}\BibitemShut {NoStop}%
\bibitem [{\citenamefont {Abbott}\ \emph {et~al.}(2019)\citenamefont {Abbott},
  \citenamefont {Allam}, \citenamefont {Andersen}, \citenamefont {Angus},
  \citenamefont {Asorey}, \citenamefont {Avelino}, \citenamefont {Avila},
  \citenamefont {Bassett}, \citenamefont {Bechtol}, \citenamefont {Bernstein}
  \emph {et~al.}}]{abbott2019first}%
  \BibitemOpen
  \bibfield  {author} {\bibinfo {author} {\bibfnamefont {T.}~\bibnamefont
  {Abbott}}, \bibinfo {author} {\bibfnamefont {S.}~\bibnamefont {Allam}},
  \bibinfo {author} {\bibfnamefont {P.}~\bibnamefont {Andersen}}, \bibinfo
  {author} {\bibfnamefont {C.}~\bibnamefont {Angus}}, \bibinfo {author}
  {\bibfnamefont {J.}~\bibnamefont {Asorey}}, \bibinfo {author} {\bibfnamefont
  {A.}~\bibnamefont {Avelino}}, \bibinfo {author} {\bibfnamefont
  {S.}~\bibnamefont {Avila}}, \bibinfo {author} {\bibfnamefont
  {B.}~\bibnamefont {Bassett}}, \bibinfo {author} {\bibfnamefont
  {K.}~\bibnamefont {Bechtol}}, \bibinfo {author} {\bibfnamefont
  {G.}~\bibnamefont {Bernstein}},  \emph {et~al.},\ }\href@noop {} {\bibfield
  {journal} {\bibinfo  {journal} {The Astrophysical Journal Letters}\ }\textbf
  {\bibinfo {volume} {872}},\ \bibinfo {pages} {L30} (\bibinfo {year}
  {2019})}\BibitemShut {NoStop}%
\bibitem [{\citenamefont {Frampton}\ \emph {et~al.}(2012)\citenamefont
  {Frampton}, \citenamefont {Ludwick},\ and\ \citenamefont
  {Scherrer}}]{frampton2012pseudo}%
  \BibitemOpen
  \bibfield  {author} {\bibinfo {author} {\bibfnamefont {P.~H.}\ \bibnamefont
  {Frampton}}, \bibinfo {author} {\bibfnamefont {K.~J.}\ \bibnamefont
  {Ludwick}}, \ and\ \bibinfo {author} {\bibfnamefont {R.~J.}\ \bibnamefont
  {Scherrer}},\ }\href@noop {} {\bibfield  {journal} {\bibinfo  {journal}
  {Physical Review D}\ }\textbf {\bibinfo {volume} {85}},\ \bibinfo {pages}
  {083001} (\bibinfo {year} {2012})}\BibitemShut {NoStop}%
\end{thebibliography}%
\end{document}